\documentclass[twocolumn,pra,superscriptaddress,longbibliography]{revtex4-1}
\usepackage{graphicx,amsmath,amssymb}
\usepackage[colorlinks=true, citecolor=blue, urlcolor=blue, linkcolor=red ]{hyperref}

\begin{document}
\title{Quantum parameter estimation in a dissipative environment}

\author{Wei Wu}

\email{wuw@lzu.edu.cn}

\affiliation{School of Physical Science and Technology, Lanzhou University,\\
 Lanzhou 730000, People's Republic of China}

\author{Chuan Shi}

\affiliation{School of Physical Science and Technology, Lanzhou University,\\
 Lanzhou 730000, People's Republic of China}

\begin{abstract}
We investigate the performance of quantum parameter estimation based on a qubit-probe in a dissipative bosonic environment beyond the traditional paradigm of weak-coupling and rotating wave approximations. By making use of an exactly numerical hierarchical equations of motion method, we analyze the influences of the non-Markovian memory effect induced by the environment and the form of probe-environment interaction on the estimation precision. It is found that (i) the non-Markovainity can effectively boost the estimation performance; and (ii) the estimation precision can be improved by introducing a perpendicular probe-environment interaction. Our results indicate the scheme of parameter estimation in a noisy environment can be optimized via engineering the decoherence mechanism.
\end{abstract}

\maketitle

\section{Introduction}\label{sec:sec1}

Ultra-sensitive parameter estimation plays an important role in both theoretical and practical researches. It has wide applications from gravitational wave detection~\cite{PhysRevLett.123.231107,PhysRevLett.123.231108}, atom clock synchronization~\cite{PhysRevLett.114.103601,PhysRevLett.113.154101}, to various high accuracy thermometries~\cite{PhysRevLett.114.220405,PhysRevB.98.045101,PhysRevX.10.011018} and magnetometers~\cite{PhysRevLett.115.200501,PhysRevLett.116.240801,Bhattacharjee_2020}. Many previous studies have revealed that certain quantum resources, for example, entanglement~\cite{Nagata726,Zou6381,PhysRevLett.121.160502,PhysRevA.92.032317} and quantum squeezing~\cite{PhysRevD.23.1693,PhysRevLett.119.193601}, can substantially improve the estimation precision and beat the shot-noise limit (standard quantum limit), which is set by the law of classical statistics. Thus, using quantum technology to attain a higher estimation accuracy has became a hot topic in the last decades, and the theory of quantum parameter estimation has been established correspondingly~\cite{RevModPhys.89.035002,RevModPhys.90.035005}. Quantum Fisher information (QFI) lies at the heart of quantum parameter estimation~\cite{683779922,683779921,Liu_2019}. Roughly speaking, it characterizes the statistical information which is extractable from a quantum state carrying the parameter of interest. In this sense, QFI theoretically determines the minimal estimation error, which is independent of specific measurement schemes. Moreover, going beyond the scope of quantum estimation theory, it has been revealed that QFI can be also used to detect the quantum phase transition of a many-body system~\cite{PhysRevA.80.012318,PhysRevA.82.022306,Wu2016,Wang_2014}, quantity the smallest evolution time for a quantum process~\cite{PhysRevA.85.052127,PhysRevLett.110.050402,Deffner_2017}, and measure the non-Markovian information flow in an open quantum system~\cite{PhysRevA.82.042103,PhysRevA.91.042110,Li_2019}.

In any practical and actual parameter estimation scheme, the quantum probe, carrying the parameter of interest, unavoidably interacts with its surrounding environment, which generally impairs the quantum resource labeling on the probe and induces the deterioration of quantum coherence. In this sense, the probe and its surrounding environment form an open quantum system~\cite{RevModPhys.59.1,RevModPhys.88.021002,RevModPhys.89.015001}, which implies the estimation performance can be severely influenced by the environment. To gain a global view and more physical insight into the quantum parameter estimation problem, the estimation scheme should be investigated within the framework of quantum dissipative dynamics and how to degrade the noise's impact should be taken into account~\cite{Demkowicz2012,PhysRevLett.113.250801,PhysRevLett.112.120405,1912.04675}. Almost all the existing studies of parameter estimation in a noisy environment restricted their attentions to some exactly solvable situations. For example, they usually assume the probe suffers a pure dephasing decoherence channel~\cite{PhysRevLett.109.233601,Razavian2019} or certain especial amplitude-damping decoherence channels~\cite{PhysRevA.88.035806,PhysRevA.87.032102,Wang_2017,PhysRevA.97.012126,SALARISEHDARAN2019126006}. Very few studies focus on the more general case, where both the dephasing mechanism and quantum relaxation are considered. Considering the fact that the real decoherence process is intricate, generalizing the study of noisy parameter estimation to a more general dissipative environment is highly desirable from both theoretical and experimental perspectives.

To address the above concern, one needs to solve the difficulty in achieving an accurate dynamical description of the quantum probe, which is coupled to a general dissipative environment. Therefore, an efficient and reliable approach is typically required. In this paper, we adopt the hierarchical equations of motion (HEOM) approach~\cite{doi:10.1143/JPSJ.58.101,YAN2004216,PhysRevE.75.031107,doi:10.1063/1.2938087,PhysRevA.85.062323} to handle this problem. The HEOM is a set of time-local differential equations for the reduced density matrix of the probe, which can provide a completely equivalent describe of the exact Schr$\ddot{\mathrm{o}}$dinger equation (or the quantum von Neumann equation). This method is beyond the usual Markovian approximation, the rotating-wave approximation (RWA), and the perturbative approximation. Thus, the HEOM can be viewed as an exactly numerical treatment of the quantum dissipative dynamics. In recent years, the HEOM approach has been successfully used to study the anomalous decoherence phenomenon in a nonlinear spin-boson model~\cite{PhysRevA.94.062116}, the quantum Zeno and anti-Zeno phenomena in a noisy environment~\cite{PhysRevA.95.042132}, as well as the influence of counter-rotating-wave terms on the measure of non-Markovianity~\cite{PhysRevA.96.032125}.

In this paper, we employ the HEOM method to study the quantum parameter estimation problem in a general dissipative environment. In Sec.~\ref{sec:sec2}, we briefly outline some basic concepts as well as the general formalism of quantum parameter estimation. In Sec.~\ref{sec:sec3}, we present three different methods employed in this paper in detail, including the HEOM approach, the general Bloch equation (GBE) technique~\cite{PhysRevB.71.035318,PhysRevB.79.125317} and the RWA treatment. Compared with the RWA approach, the effect of counter-rotating-wave terms is considered in the GBE method. Thus it can be employed as a benchmark of the purely numerical HEOM approach. The main results and the conclusions of this paper are drawn in Sec.~\ref{sec:sec4} and Sec.~\ref{sec:sec5}, respectively. Throughout the paper, we set $\hbar=k_{\mathrm{B}}=1$ for the sake of simplicity, and all the other units are dimensionless as well.

\section{Noisy quantum parameter estimation}\label{sec:sec2}

In the theory of quantum parameter estimation, the parameter's information is commonly encoded into the state of the quantum probe via a unitary~\cite{PhysRevLett.79.3865,Hauke2016,McCormick2019} or non-unitary dynamics~\cite{PhysRevLett.109.233601,Razavian2019,PhysRevA.88.035806,PhysRevA.87.032102,Wang_2017,PhysRevA.97.012126,SALARISEHDARAN2019126006,Haase_2018,PhysRevLett.123.040402,tamascelli2020quantum}. Then, one can extract the message of the parameter $\theta$ from the output state of the probe $\rho_{\theta}$ via repeated quantum measurements. In such quantum parameter estimation process, one can not completely eliminate all the errors and estimate $\theta$ precisely. There exists a minimal estimation error, which can not be removed by optimizing the estimation scheme, is given by the famous quantum Cram\'{e}r-Rao bound~\cite{683779922,683779921,Liu_2019}
\begin{equation}\label{eq:eq1}
\delta \theta\geq\frac{1}{\sqrt{\upsilon F(\theta)}},
\end{equation}
where $\delta \theta$ the root mean square of $\theta$, $\upsilon$ is the number of repeated measurements (in this paper, we set $\upsilon=1$ for the sake of convenience), and $F(\theta)\equiv\mathrm{Tr}(\rho_{\theta}\hat{L}_{\theta})$ with $\hat{L}_{\theta}$ determined by $\partial_{\theta}\rho_{\theta}=\frac{1}{2}(\hat{L}_{\theta}\rho_{\theta}+\rho_{\theta}\hat{L}_{\theta})$ is the QFI with respect to the output state $\rho_{\theta}$. From Eq.~(\ref{eq:eq1}), one can immediately find that the optimal estimation precision is completely decided by the value of QFI: the larger the QFI, the smaller the estimation error is. How to saturate the smallest theoretical error (or boost the QFI) is the most crucial problem in the field of quantum parameter estimation.

To compute the QFI from the $\theta$-dependent density operator $\rho_{\theta}$, one first needs to diagonalize $\rho_{\theta}$ as $\rho_{\theta}=\sum_{\ell}\xi_\ell|\xi_\ell\rangle\langle\xi_\ell|$, where $\xi_{\ell}\equiv\xi_{\ell}(\theta)$ and $|\xi_{\ell}\rangle\equiv|\xi_{\ell}(\theta)\rangle$ are eigenvalues and eigenvectors of $\rho_{\theta}$, respectively. Then, the QFI can be computed as~\cite{Liu_2019}
\begin{equation}\label{eq:eq2}
\begin{split}
F(\theta)=&\sum_{\ell}\frac{(\partial_{\theta}\xi_{\ell})^{2}}{\xi_{\ell}}+\sum_{\ell}4\xi_{\ell}\langle\partial_{\theta}\xi_{\ell}|\partial_{\theta}\xi_{\ell}\rangle\\
&-\sum_{\ell,\ell'}\frac{8\xi_{\ell}\xi_{\ell'}}{\xi_{\ell}+\xi_{\ell'}}|\langle\partial_{\theta}\xi_{\ell}|\xi_{\ell'}\rangle|^{2}.
\end{split}
\end{equation}
Specially, for a two-dimensional density operator described in the Bloch representation, namely, $\rho_{\theta}=\frac{1}{2}(\mathbf{1}_{2}+\langle\pmb{\hat{\underline{\sigma}}}\rangle\cdot\pmb{\hat{\sigma}})$ with $\langle\pmb{\hat{\underline{\sigma}}}\rangle\equiv(\langle\hat{\sigma}_{x}\rangle,\langle\hat{\sigma}_{y}\rangle,\langle\hat{\sigma}_{z}\rangle)^{\mathbb{T}}$ being the Bloch vector and $\pmb{\hat{\sigma}}\equiv(\hat{\sigma}_{x},\hat{\sigma}_{y},\hat{\sigma}_{z})$ being the vector of Pauli matrices, Eq.~(\ref{eq:eq2}) can be further simplified to~\cite{Liu_2019}
\begin{equation}\label{eq:eq3}
F(\theta)=|\partial_{\theta}\langle\pmb{\hat{\underline{\sigma}}}\rangle|^{2}+\frac{(\langle\pmb{\hat{\underline{\sigma}}}\rangle\cdot\partial_{\theta}\langle\pmb{\hat{\underline{\sigma}}}\rangle)^{2}}{1-|\langle\pmb{\hat{\underline{\sigma}}}\rangle|^{2}}.
\end{equation}
For pure state case, the above equation reduces to $F(\theta)=|\partial_{\theta}\langle\pmb{\hat{\underline{\sigma}}}\rangle|^{2}$. Compared with Eq.~(\ref{eq:eq2}), Eq.~(\ref{eq:eq3}) is more computable in practice, because it avoids the operation of diagonalization.

In this work, we assume a qubit, acting as the probe and carrying the parameter of interest, is linearly coupled to a dissipative environment. The Hamiltonian of the quantum probe is described by $\hat{H}_{\mathrm{s}}=\frac{1}{2}\Delta\hat{\sigma}_{x}$, where $\Delta$ represents the frequency of tunneling between the two levels of the qubit and \emph{is the encoded parameter to be estimated in this paper}. We assume the dissipative environment is stimulated by a set of harmonic oscillators, i.e., $\hat{H}_{\mathrm{b}}=\sum_{k}\omega_{k}\hat{b}_{k}^{\dagger}\hat{b}_{k}$, where $\hat{b}_{k}^{\dagger}$ and $\hat{b}_{k}$ are the creation and annihilation operators of the $k$th harmonic oscillator with corresponding frequency $\omega_{k}$, respectively. Thus, the Hamiltonian of the whole qubit-probe plus the environment is given by $\hat{H}=\hat{H}_{\mathrm{s}}+\hat{H}_{\mathrm{b}}+\hat{H}_{\mathrm{i}}$. Here, we assume the probe-environment interaction part can be described in the following linear form
\begin{equation}\label{eq:eq4}
\hat{H}_{\mathrm{i}}=\mathcal{\hat{S}}\otimes\mathcal{\hat{B}},
\end{equation}
where $\hat{\mathcal{S}}$ denotes the probe's operator coupled to its surrounding environment, and $\mathcal{\hat{B}}\equiv\sum_{k}g_{k}(\hat{b}_{k}^{\dagger}+\hat{b}_{k})$ with $g_{k}$ being the coupling strength between the probe and the $k$th environmental mode.

After a period of non-unitary dynamics, the information of $\Delta$ is then encoded in the reduced density operator of the probe, namely, $\varrho_{\mathrm{s}}(t)\equiv \mathrm{Tr}_{\mathrm{b}}[e^{-i\hat{H}t}\varrho_{\mathrm{sb}}(0)e^{i\hat{H}t}]$. Here, $\varrho_{\mathrm{sb}}(0)$ is the initial state of the whole Hamiltonian. Generally speaking, the ultimate estimation precision associated with $\varrho_{\mathrm{s}}(t)$ depends a number of factors. In this paper, we concentrate on the following two elements: \emph{the characteristic of the environment} and \emph{the form of the probe-environment coupling operator}. The property of the environment is mainly reflected by environmental auto-correlation function, which is defined by
\begin{equation}\label{eq:eq5}
\alpha(t)\equiv\mathrm{Tr}_{\mathrm{b}}\Big{(}e^{it\hat{H}_{\mathrm{b}}}\hat{\mathcal{B}}e^{-it\hat{H}_{\mathrm{b}}}\hat{\mathcal{B}}\varrho_{\mathrm{b}}\Big{)},
\end{equation}
with $\varrho_{\mathrm{b}}$ being the initial state of the environment. Thus, we shall discuss the effect of $\alpha(t)$ and $\mathcal{\hat{S}}$ on the QFI with respect to $\varrho_{\mathrm{s}}(t)$. The determination of the QFI requires the knowledge of the reduced density operator $\varrho_{\mathrm{s}}(t)$. Unfortunately, except in a few special situations, the exact expression of $\varrho_{\mathrm{s}}(t)$ is generally difficult to obtain. To overcome this difficulty, we would like to adopt the following three different methods to evaluate $F(\Delta)$.

\section{Methodology}\label{sec:sec3}

In this section, we introduce the dynamical formulations employed in our study. The first one is the HEOM method, which can provide rigorous numerical results. As comparisons, we also present two analytical methods: the GBE and the RWA approaches. In this paper, we assume the initial state of the whole probe-environment system has a factorizing form, i.e., $\varrho_{\mathrm{sb}}(0)=\varrho_{\mathrm{s}}(0)\otimes\varrho_{\mathrm{b}}$, where $\varrho_{\mathrm{b}}=|\mathbf{0}_{k}\rangle\langle \mathbf{0}_{k}|$ with $|\mathbf{0}_{k}\rangle\equiv\bigotimes_{k}|0_{k}\rangle$ being the Fock vacuum state of the environment.

\subsection{HEOM}

The HEOM can be viewed as a bridge linking the well-known Schr$\ddot{\mathrm{o}}$dinger equation, which is exact but generally difficult to solve straightforwardly, and a set of ordinary differential equations, which can be handled numerically by using the Runge-Kutta method. How to establish such a connection, which should be elaborately designed and avoid losing any important dynamical feature of the the quantum probe, is the most important step in the HEOM treatment~\cite{PhysRevA.98.012110,PhysRevA.98.032116}. In many previous references, the HEOM algorithm is realized by making use of the path-integral influence functional approach~\cite{PhysRevE.75.031107,doi:10.1063/1.2938087}. In this paper, we establish the HEOM in an alternative way: within the framework of the non-Markovian quantum state diffusion approach~\cite{PhysRevA.98.012110,PhysRevLett.113.150403,Suess2015}.

The dynamics of $\hat{H}$ is determined by the Schr$\ddot{\mathrm{o}}$dinger equation $\partial_{t}|\Psi_{\mathrm{sb}}(t)\rangle=-i\hat{H}|\Psi_{\mathrm{sb}}(t)\rangle$, where $|\Psi_{\mathrm{sb}}(t)\rangle$ is the pure-state wave function of the whole probe-environment system. Any straightforward treatment of the above Schr$\ddot{\mathrm{o}}$dinger equation can be rather troublesome, because of the large number of degrees of freedom. However, by employing the bosonic coherent state, which is defined by $|\mathbf{z}\rangle=\bigotimes_{k}|z_{k}\rangle$ with $|z_{k}\rangle\equiv e^{z_{k}\hat{b}_{k}^{\dagger}}|0_{k}\rangle$, one can recast the original Schr$\ddot{\mathrm{o}}$dinger equation into the following stochastic quantum state diffusion equation (see Appendix for more details)
\begin{equation}\label{eq:eq6}
\begin{split}
\frac{\partial}{\partial t}|\psi_{t}(\textbf{z}^{*})\rangle=&-i\hat{H}_{\mathrm{s}}|\psi_{t}(\textbf{z}^{*})\rangle+\hat{\mathcal{S}}\textbf{z}_{t}^{*}|\psi_{t}(\textbf{z}^{*})\rangle\\
&-\hat{\mathcal{S}}\int_{0}^{t}d\tau\alpha(t-\tau)\frac{\delta}{\delta \textbf{z}_{\tau}^{*}}|\psi_{t}(\textbf{z}^{*})\rangle,
\end{split}
\end{equation}
where $|\psi_{t}(\textbf{z}^{*})\rangle\equiv\langle \textbf{z}|\Psi_{\mathrm{sb}}(t)\rangle$ is the total pure-state wave function in the coherent-state representation, the variable $\textbf{z}_{t}\equiv i\sum_{k}g_{k}z_{k}e^{-i\omega_{k}t}$ can be regarded as a stochastic Gaussian colored noise satisfying $\mathcal{M}\{\textbf{z}_{t}\}=\mathcal{M}\{\textbf{z}_{t}^{*}\}=0$ and $\mathcal{M}\{\textbf{z}_{t}\textbf{z}_{\tau}^{*}\}=\alpha(t-\tau)$. Here $\mathcal{M}\{...\}$ denotes the statistical mean over all the possible quantum trajectories, and $\alpha(t)=\sum_{k}g_{k}^{2}e^{-i\omega_{k}t}$ is the auto-correlation function at zero temperature. In this paper, we concentrate on an Ornstein-Uhlenbeck-type auto-correlation function, namely
\begin{equation}\label{eq:eq7}
\alpha(t)=\frac{1}{2}\Gamma\gamma e^{-\gamma t},
\end{equation}
where $\Gamma$ can be viewed as the probe-environment coupling strength and $\gamma$ is connected to the memory time of the environment.

Notice that the auto-correlation function has an exponential form of time, which means $\partial_{t}\alpha(t)=-\gamma\alpha(t)$. Using this property, we can replace the stochastic quantum state diffusion equation in Eq.~(\ref{eq:eq6}) with a set of hierarchial equations of the pure-state wave function $|\psi_{t}(\textbf{z}^{*})\rangle$ as follows~\cite{PhysRevA.98.012110,PhysRevLett.113.150403,Suess2015}
\begin{equation}\label{eq:eq8}
\begin{split}
\frac{\partial}{\partial t}|\psi_{t}^{(m)}\rangle=&(-i\hat{H}_{\mathrm{s}}-m\gamma+\hat{\mathcal{S}}\textbf{z}_{t}^{*})|\psi_{t}^{(m)}\rangle\\
&+\frac{1}{2}m\Gamma\gamma\hat{\mathcal{S}}|\psi_{t}^{(m-1)}\rangle-\hat{\mathcal{S}}|\psi_{t}^{(m+1)}\rangle,
\end{split}
\end{equation}
where
\begin{equation*}
|\psi_{t}^{(m)}\rangle\equiv\Bigg{[}\int_{0}^{t}d\tau C(t-\tau)\frac{\delta}{\delta \textbf{z}_{\tau}^{*}}\Bigg{]}^{m}|\psi_{t}(\textbf{z}^{*})\rangle,
\end{equation*}
are auxiliary pure-state wave functions. The hierarchy equation of $|\psi_{t}(\textbf{z}^{*})\rangle$ in Eq.~(\ref{eq:eq8}) no longer contains functional derivatives, but still has stochastic noise terms which hinder the efficiency of numerical simulation. To extract a deterministic equation of motion for the reduced density operator, one needs to trace out the degrees of freedom of the environment by taking the statistical mean over all the possible quantum trajectories~\cite{PhysRevA.98.012110,PhysRevA.98.032116}. The expression of the reduced density operator is then given by $\varrho_{\mathrm{s}}(t)=\varrho_{t}\equiv\mathcal{M}\big{\{}|\psi_{t}(\textbf{z}^{*})\rangle\langle\psi_{t}(\textbf{z}^{*})|\big{\}}$. As shown in Ref.~\cite{PhysRevA.98.012110} show, the equation of motion for $\varrho_{t}$ can be derived from Eq.~(\ref{eq:eq8}), and reads
\begin{equation}\label{eq:eq9}
\begin{split}
\frac{d}{dt}\varrho_{t}^{(m,n)}=&-i\Big{[}\hat{H}_{\mathrm{s}},\varrho_{t}^{(m,n)}\Big{]}-\gamma(m+n)\varrho_{t}^{(m,n)}\\
&+\frac{1}{2}\Gamma\gamma\Big{[}m\hat{\mathcal{S}}\varrho_{t}^{(m-1,n)}+n\varrho_{t}^{(m,n-1)}\hat{\mathcal{S}}\Big{]}\\
&-\Big{[}\hat{\mathcal{S}},\varrho_{t}^{(m+1,n)}\Big{]}+\Big{[}\hat{\mathcal{S}},\varrho_{t}^{(m,n+1)}\Big{]},
\end{split}
\end{equation}
where $\varrho_{t}^{(m,n)}\equiv\mathcal{M}\big{\{}|\psi_{t}^{(m)}(\textbf{z}^{*})\rangle\langle\psi_{t}^{(n)}(\textbf{z}^{*})|\big{\}}$ are auxiliary reduced density operators. Eq.~(\ref{eq:eq9}) is nothing but a set of ordinary differential equations, which shall be handled in our numerical simulations.

The initial state condition of the auxiliary operators are $\varrho^{(0,0)}_{t}=\varrho_{\mathrm{s}}(0)$ and $\varrho^{(m>0,n>0)}_{t}=0$. In numerical simulations, we need to truncate the hierarchical equations by choosing a sufficiently large integer $N$. All the terms of $\varrho^{(m,n)}_{t}$ with $m+n>N$ are set to zero, while the terms of $\varrho^{(m,n)}_{t}$ with $m+n\leq N$ consist of a closed set of ordinary differential equations which can be solved directly by using the fourth-order Runge-Kutta method. It is necessary to emphasize that no approximation is invoked in the above derivation from Eq.~(\ref{eq:eq6}) to Eq.~(\ref{eq:eq9}), which means the mapping from the original Schr$\ddot{\mathrm{o}}$dinger equation to the hierarchy equations given by Eq.~(\ref{eq:eq9}) is exact. In this sense, the numerics obtained from Eq.~(\ref{eq:eq9}) should be viewed as rigorous results.

\subsection{GBE}

If $\hat{\mathcal{S}}=\hat{\sigma}_{z}$, Eq.~(\ref{eq:eq4}) has a purely transversal or perpendicular interaction (recalling that $\hat{H}_{\mathrm{s}}=\frac{1}{2}\Delta\hat{\sigma}_{x}$). Thus, the probe-environment system has the same structure as the famous spin-boson model, which leads to both the loss of information and the dissipation of energy. The equation of motion of the spin-boson model is governed by the quantum von Neumann equation $\partial_{t}\varrho_{\mathrm{sb}}(t)=-i[\hat{H},\varrho_{\mathrm{sb}}(t)]$, which provides an exact dynamical prediction. Applying the Zwanzig's projection technique with the Born approximation, the quantum von Neumann equation can be transformed to the well-known Zwanzing-Nakajima master equation~\cite{10.1143/PTP.20.948,doi:10.1063/1.1731409}
\begin{equation}\label{eq:eq10}
\frac{\partial}{\partial t}\varrho_{\mathrm{s}}(t)=-i\mathcal{\hat{L}}_{\mathrm{s}}\varrho_{\mathrm{s}}(t)-\int_{0}^{t}d\tau\hat{\Sigma}(t-\tau)\varrho_{\mathrm{s}}(\tau),
\end{equation}
here $\hat{\Sigma}(t)$ is the self-energy super-operator
\begin{equation}\label{eq:eq11}
\hat{\Sigma}(t)=\mathrm{Tr}_{\mathrm{b}}\Big{[}\mathcal{\hat{L}}_{\mathrm{i}}e^{-it\mathcal{\hat{Q}}(\mathcal{\hat{L}}_{\mathrm{s}}+\mathcal{\hat{L}}_{\mathrm{b}}+\mathcal{\hat{L}}_{\mathrm{i}})}\mathcal{\hat{L}}_{\mathrm{i}}\varrho_{\mathrm{b}}\Big{]},
\end{equation}
where $\mathcal{\hat{L}}_{\mathrm{x}}$ with $\mathrm{x}=\mathrm{s},\mathrm{b},\mathrm{i}$ is the Liouvillian superoperator satisfying  $\mathcal{\hat{L}}_{\mathrm{x}}\mathcal{\hat{O}}=[\hat{H}_{\mathrm{x}},\mathcal{\hat{O}}]$, and $\mathcal{\hat{Q}}\equiv 1-\varrho_{\mathrm{b}}\mathrm{Tr}_{\mathrm{b}}$ is Zwanzig's projection superoperator. From Eq.~(\ref{eq:eq10}), one can notice the evolution of $\varrho_{\mathrm{s}}(t)$ depends on $\varrho_{\mathrm{s}}(\tau)$ at all the earlier times $0<\tau<t$, implying the memory effect from the environment has been considered and is incorporated into the self-energy super-operator $\hat{\Sigma}(t-\tau)$. Thus, the result from Eq.~(\ref{eq:eq9}) is then non-Markovian.

The exact treatment of the above Zwanzing-Nakajima master equation in Eq.~(\ref{eq:eq10}) is challenging. Fortunately, the self-energy super-operator of Eq.~(\ref{eq:eq11}) can be expanded in powers of the interaction Liouvillian $\mathcal{\hat{L}}_{\mathrm{i}}$. Only retaining the lowest-order term in the series, $\hat{\Sigma}(t)$ can be approximated as~\cite{PhysRevB.71.035318,PhysRevB.79.125317,WU2020168203}
\begin{equation}\label{eq:eq12}
\hat{\Sigma}(t)\simeq\mathrm{Tr}_{\mathrm{b}}\Big{[}\mathcal{\hat{L}}_{\mathrm{i}}e^{-it(\mathcal{\hat{L}}_{\mathrm{s}}+\mathcal{\hat{L}}_{\mathrm{b}})}\mathcal{\hat{L}}_{\mathrm{i}}\varrho_{\mathrm{b}}\Big{]}.
\end{equation}
Eq.~(\ref{eq:eq10}) together with the approximate $\hat{\Sigma}(t)$ in Eq.~(\ref{eq:eq12}) constitute a general non-Markovian quantum master equation, which has been widely used in many previous studies~\cite{PhysRevB.71.035318,PhysRevB.79.125317,PhysRevA.89.062113}.

By introducing the time-dependent Bloch vector $\langle\pmb{\hat{\underline{\sigma}}}(t)\rangle$ with $\langle\hat{\sigma}_{i}(t)\rangle\equiv\mathrm{Tr}_{\mathrm{s}}[\hat{\sigma}_{i}\varrho_{\mathrm{s}}(t)]$, one can rewrite the above general quantum master equation as the following GBE~\cite{PhysRevB.71.035318,PhysRevB.79.125317}
\begin{equation}\label{eq:eq13}
\frac{d}{dt}\langle\pmb{\hat{\underline{\sigma}}}(t)\rangle=\mathfrak{\hat{T}}(t)\diamond\langle\pmb{\hat{\underline{\sigma}}}(t)\rangle,
\end{equation}
where $\diamond$ denotes the convolution and
\begin{equation*}
\mathfrak{\hat{T}}(t)=\left[
                       \begin{array}{ccc}
                         -\mathfrak{A}(t) & 0 & 0 \\
                         0 & -\mathfrak{B}(t) & -\Delta\delta(t) \\
                         0 & \Delta\delta(t) & 0 \\
                       \end{array}
                     \right],
\end{equation*}
with $\mathfrak{A}(t)=4\cos(\Delta t)\alpha(t)$, $\mathfrak{B}(t)=4\alpha(t)$. By means of the Laplace transform, one can find
\begin{equation}\label{eq:eq14}
\begin{split}
\langle\hat{\sigma}_{i}(\lambda)\rangle\equiv&\int_{0}^{\infty}dt\langle\hat{\sigma}_{i}(t)\rangle e^{-\lambda t}\\
=&\sum_{j}\mathfrak{F}_{ij}(\lambda)\langle\hat{\sigma}_{j}(0)\rangle,
\end{split}
\end{equation}
where $i,j=x,y,z$. For the perpendicular probe-environment interaction case, the non-vanishing terms of $\mathfrak{F}_{ij}(\lambda)$ are
\begin{equation*}
\mathfrak{F}_{xx}(\lambda)=[\lambda+\mathfrak{A}(\lambda)]^{-1},
\end{equation*}
\begin{equation*}
\mathfrak{F}_{yy}(\lambda)=\bigg{[}\lambda+\mathfrak{B}(\lambda)+\frac{\Delta^{2}}{\lambda}\bigg{]}^{-1},
\end{equation*}
\begin{equation*}
\mathfrak{F}_{zz}(\lambda)=\lambda^{-1}[\lambda+\mathfrak{B}(\lambda)]\mathfrak{F}_{yy}(\lambda),
\end{equation*}
\begin{equation*}
\mathfrak{F}_{yz}(\lambda)=-\mathfrak{F}_{zy}(\lambda)=-\Delta\lambda^{-1}\mathfrak{F}_{yy}(\lambda).
\end{equation*}
Then, for an arbitrary given initial state $\langle\pmb{\hat{\underline{\sigma}}}(0)\rangle$, the dynamics of $\langle\pmb{\hat{\underline{\sigma}}}(t)\rangle$ can be completely determined by the GBE method in Eq.~(\ref{eq:eq14}) with the help of inverse Laplace transform.

\subsection{RWA}

For the purely perpendicular interaction case, one can use an alternative method, the RWA approach, to obtain the dynamical behavior of the probe. The RWA can remove the counter-rotating-wave terms in $\hat{H}$ and obtain the following approximate Hamiltonian
\begin{equation}\label{eq:eq15}
\hat{H}_{\mathrm{RWA}}=\frac{\Delta}{2}\hat{\sigma}_{x}+\sum_{k}\omega_{k}\hat{b}_{k}^{\dagger}\hat{b}_{k}+\sum_{k}g_{k}\big{(}\hat{\sigma}_{-}\hat{b}_{k}^{\dagger}+\hat{\sigma}_{+}\hat{b}_{k}\big{)},
\end{equation}
where $\hat{\sigma}_{+}\equiv|+\rangle\langle -|$ and $\hat{\sigma}_{-}\equiv|-\rangle\langle +|$ with $|\pm\rangle$ being the eigenvectors of $\hat{\sigma}_{x}$, i.e., $\hat{\sigma}_{x}|\pm\rangle=\pm|\pm\rangle$. The Hamiltonian $\hat{H}_{\mathrm{RWA}}$ commutes with the total excitation number operator $\hat{\mathcal{N}}=\hat{\sigma}_{+}\hat{\sigma}_{-}+\sum_{k}\hat{b}_{k}^{\dagger}\hat{b}_{k}$, which is thus a constant of motion and can greatly simplify the reduced dynamical solution of the probe in this situation.

At zero temperature, the reduced dynamics of the probe is exactly solvable in the RWA case and can be conveniently expressed in the basis of $\{|+\rangle,|-\rangle\}$ as follows
\begin{equation}\label{eq:eq16}
\varrho_{\mathrm{s}}(t)=\left[
          \begin{array}{cc}
            \varrho_{++}(0)\mathcal{G}_{t}^{2} & \varrho_{+-}(0)\mathcal{G}_{t}e^{-i\Delta t} \\
            \varrho_{-+}(0)\mathcal{G}_{t}e^{i\Delta t} & 1-\varrho_{++}(0)\mathcal{G}_{t}^{2} \\
          \end{array}
        \right],
\end{equation}
where $\mathcal{G}_{t}$ is the the decay factor. For the Ornstein-Uhlenbeck-type auto-correlation function considered in this paper, the exact expression of $\mathcal{G}_{t}$ is given by~\cite{PhysRevA.96.032125}
\begin{equation}\label{eq:eq17}
\begin{split}
\mathcal{G}_{t}=&\exp\Big{(}-\frac{1}{2}\gamma t\Big{)}\bigg{[}\cosh\Big{(}\frac{1}{2}\Omega t\Big{)}+\frac{\gamma}{\Omega}\sinh\Big{(}\frac{1}{2}\Omega t\Big{)}\bigg{]},
\end{split}
\end{equation}
with $\Omega\equiv\sqrt{\gamma^{2}-2\gamma\Gamma}$. As showed in many previous studies~\cite{PhysRevA.85.062323,PhysRevA.96.032125}, the RWA is acceptable in the weak probe-environment coupling regime, we thus expect it can provide a reasonable prediction in the above region.

\subsection{Comparison}\label{subsec:subsec3d}

\begin{figure}
\centering
\includegraphics[angle=0,width=4.25cm]{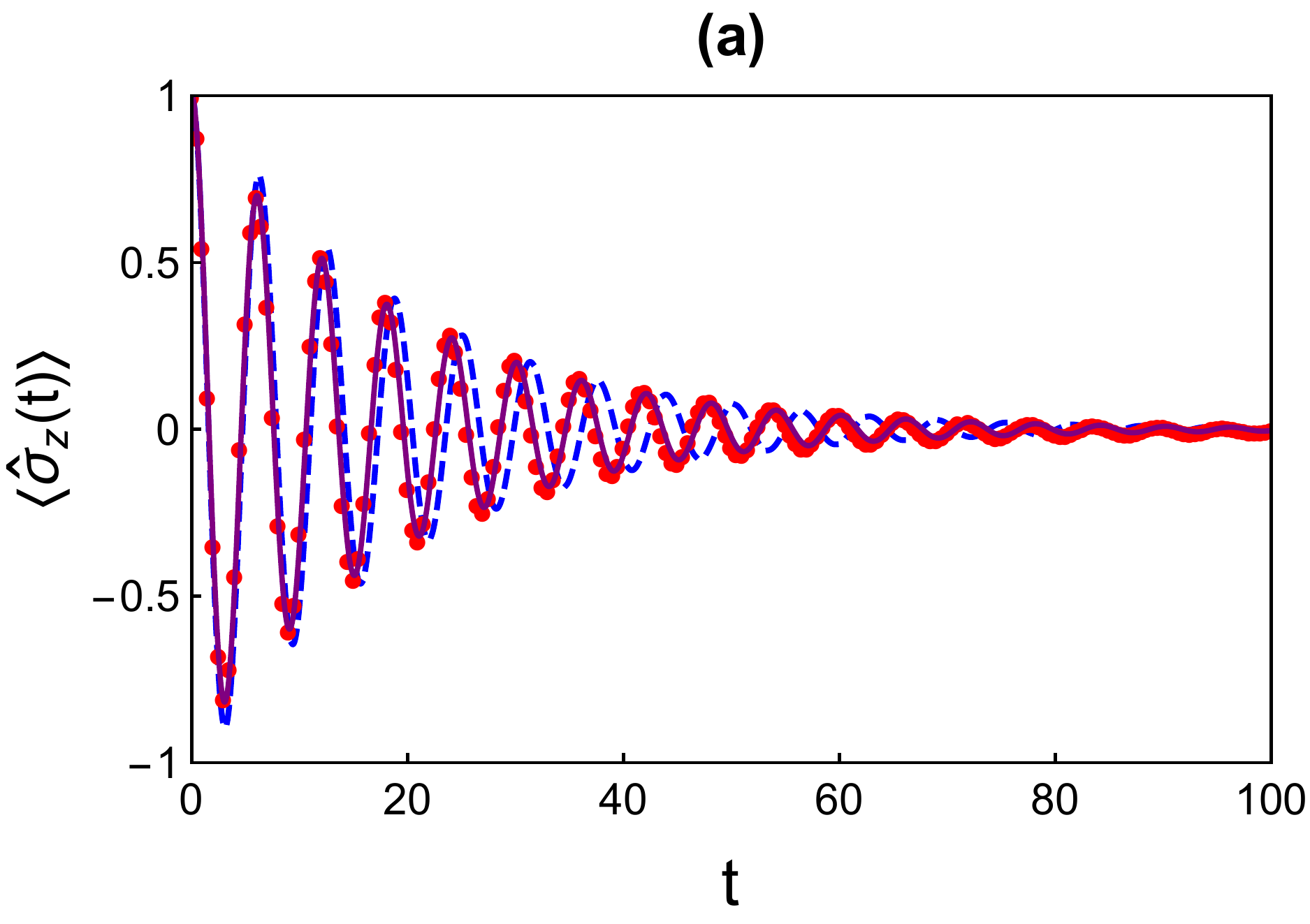}
\includegraphics[angle=0,width=4.25cm]{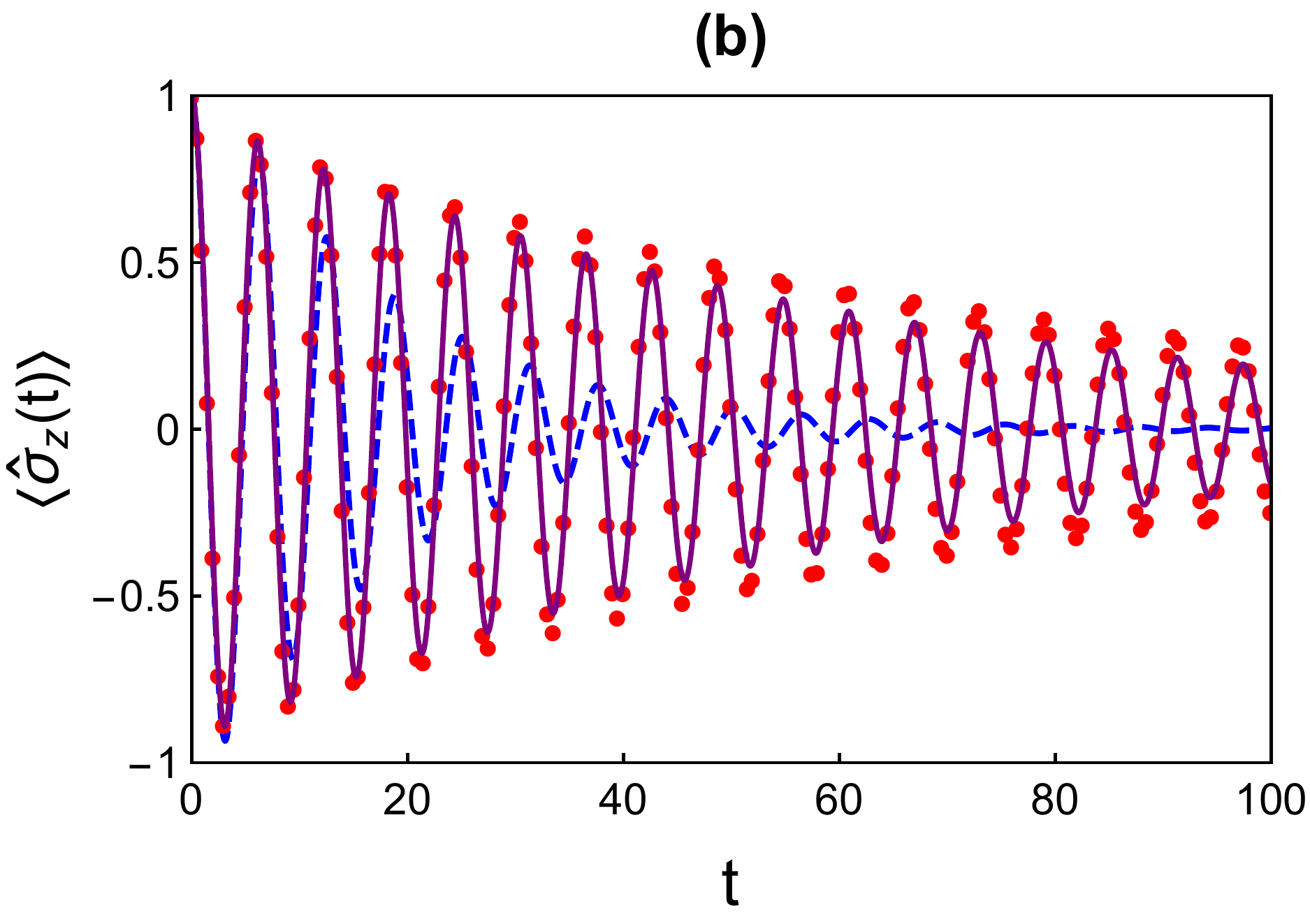}
\includegraphics[angle=0,width=4.25cm]{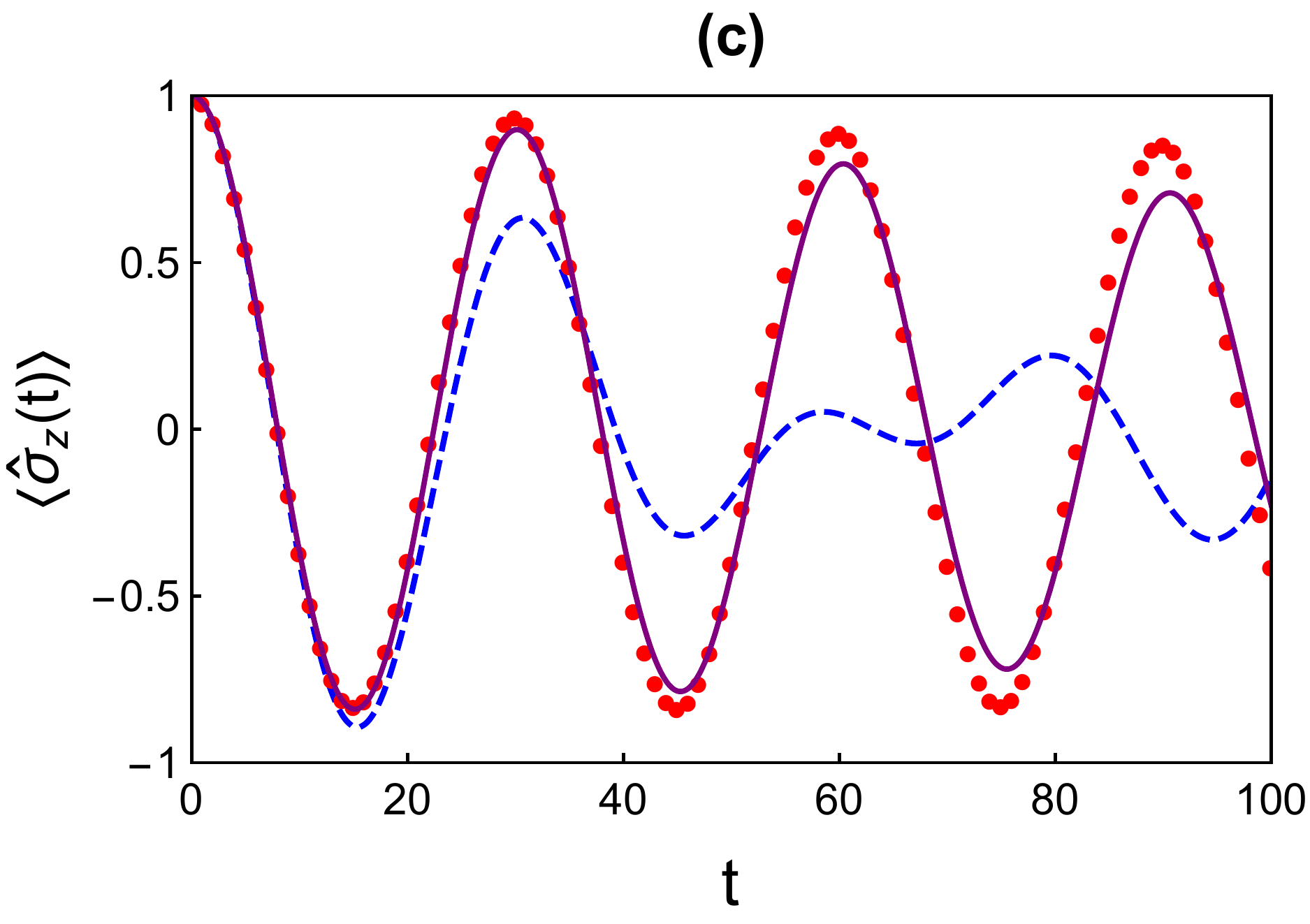}
\includegraphics[angle=0,width=4.25cm]{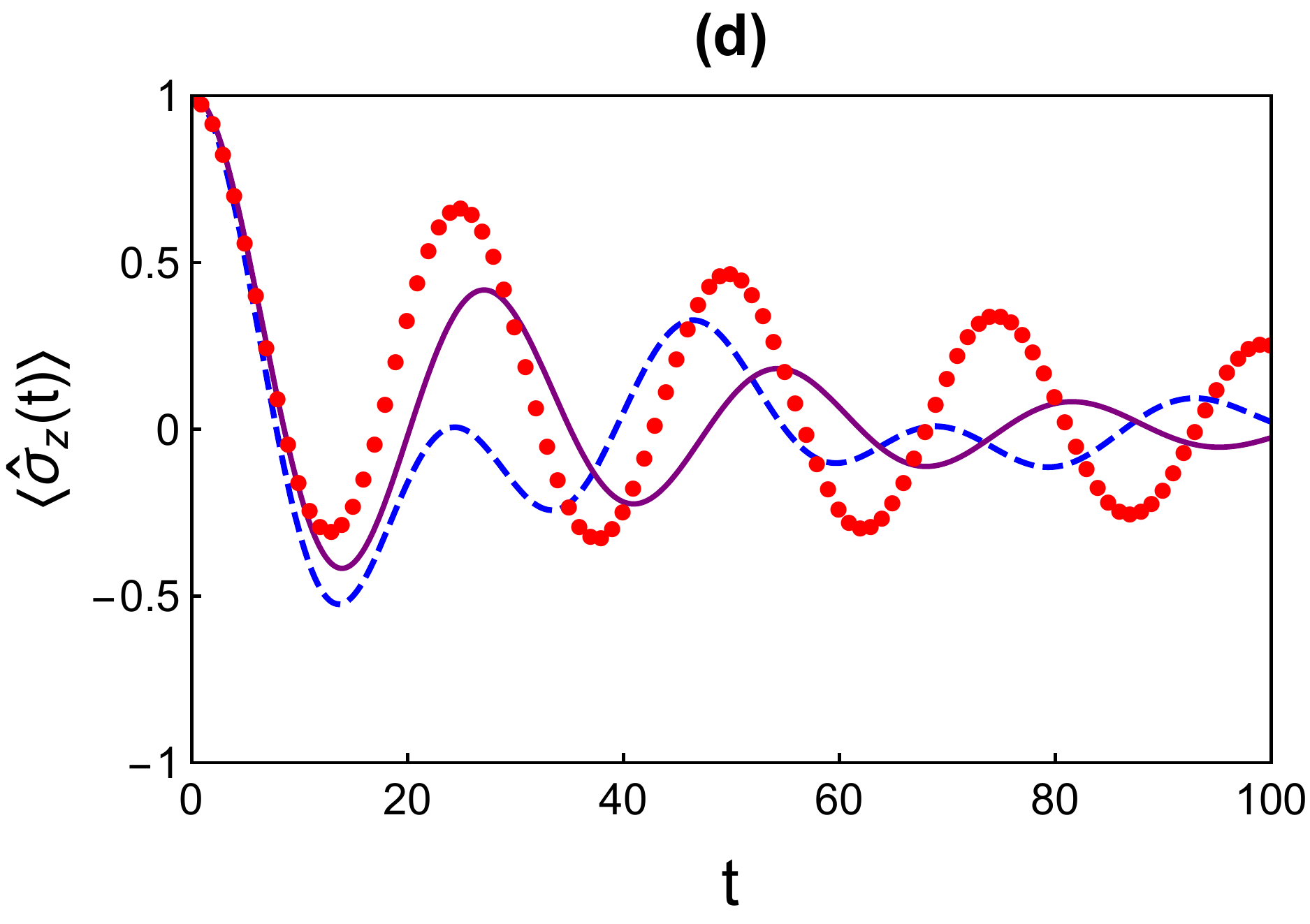}
\caption{The dynamics of the population difference $\langle\hat{\sigma}_{z}(t)\rangle$ with $\langle\hat{\sigma}_{z}(0)\rangle=1$. The purple solid lines are exact numerical simulations from the HEOM method, the red circles are results from the GBE approach and the blue dashed lines are obtained under the assumption of RWA. Parameters are chosen as: (a) $\gamma=10\Gamma$, $\Gamma=0.1$ and $\Delta=1$; (b) $\gamma=4\Gamma$, $\Gamma=0.1$ and $\Delta=1$; (c) $\gamma=0.2\Gamma$, $\Gamma=0.1$ and $\Delta=0.2$; (d) $\gamma=0.2\Gamma$, $\Gamma=0.25$ and $\Delta=0.2$. }\label{fig:fig1}
\end{figure}

In Fig.~\ref{fig:fig1}, we display the dynamics of the population difference $\langle\hat{\sigma}_{z}(t)\rangle$ of the qubit-probe, which is a very common quantity of interest in experiments. For the RWA case, the exact expression of the population difference $\langle\hat{\sigma}_{z}(t)\rangle$ is given by $\langle\hat{\sigma}_{z}(t)\rangle_{\mathrm{RWA}}=\mathcal{G}_{t}\cos(\Delta t)$.

For the Ornstein-Uhlenbeck-type correlation function considered in this paper, the boundary between Markovian and non-Markovian
regimes can be approximately specified by the ratio of $\gamma/\Gamma$~\cite{PhysRevA.95.042132,PhysRevLett.99.160502}. When $\gamma/\Gamma$ is large, the correlation function reduces to a delta correlated auto-correlation function, i.e., $\alpha(t-\tau)\simeq\Gamma\delta(t-\tau)$, which means the environment is memoryless and the decoherence dynamics is Markovian. On the contrary, if $\gamma/\Gamma$ is small, the environmental memory effect can not be neglected and the corresponding decoherence is then non-Markovian. In fact, when $\gamma/\Gamma\rightarrow\infty$, one can demonstrate that the hierarchical equations in Eq.~(\ref{eq:eq8}) can reduce to the common Markovian Lindblad-type master equation by only considering the zeroth order of the terminator~\cite{PhysRevLett.113.150403}. The relation between $\gamma/\Gamma$ and the degree of non-Markovianity has been studied by making use of trace distance~\cite{PhysRevA.81.062124,Wu2018} and dynamical divisibility~\cite{PhysRevA.83.062115}, these studies are consistent with our above analysis.

We first consider the Markovian case, say $\lambda/\Gamma=10$ in Fig.~\ref{fig:fig1}(a). A good agreement is found between results from the HEOM and the GBE, while the prediction from the RWA exhibits a small deviation from the above two approaches. Such deviation disappears if the probe-environment coupling becomes further weaker. Thus, three different approaches present a consistent result in Markovian and weak-coupling regime. In the non-Markovian regime, the result from the GBE can still be in qualitative agreement with that of the numerical HEOM method if the coupling strength is weak, see Fig.~\ref{fig:fig1} (c). However, when the coupling becomes stronger, such as the parameters chosen in Fig.~\ref{fig:fig1} (d), the GBE exhibits a relatively large deviation compared with the result from HEOM, probably because it neglects the higher-order terms of the probe-environment coupling. On the contrary, the result calculated with RWA gives a qualitatively incorrect conclusion in the entire non-Markovian regime, unless one only focuses on the short-time behavior of the population difference.

\begin{figure*}
\centering
\includegraphics[angle=0,width=5.5cm]{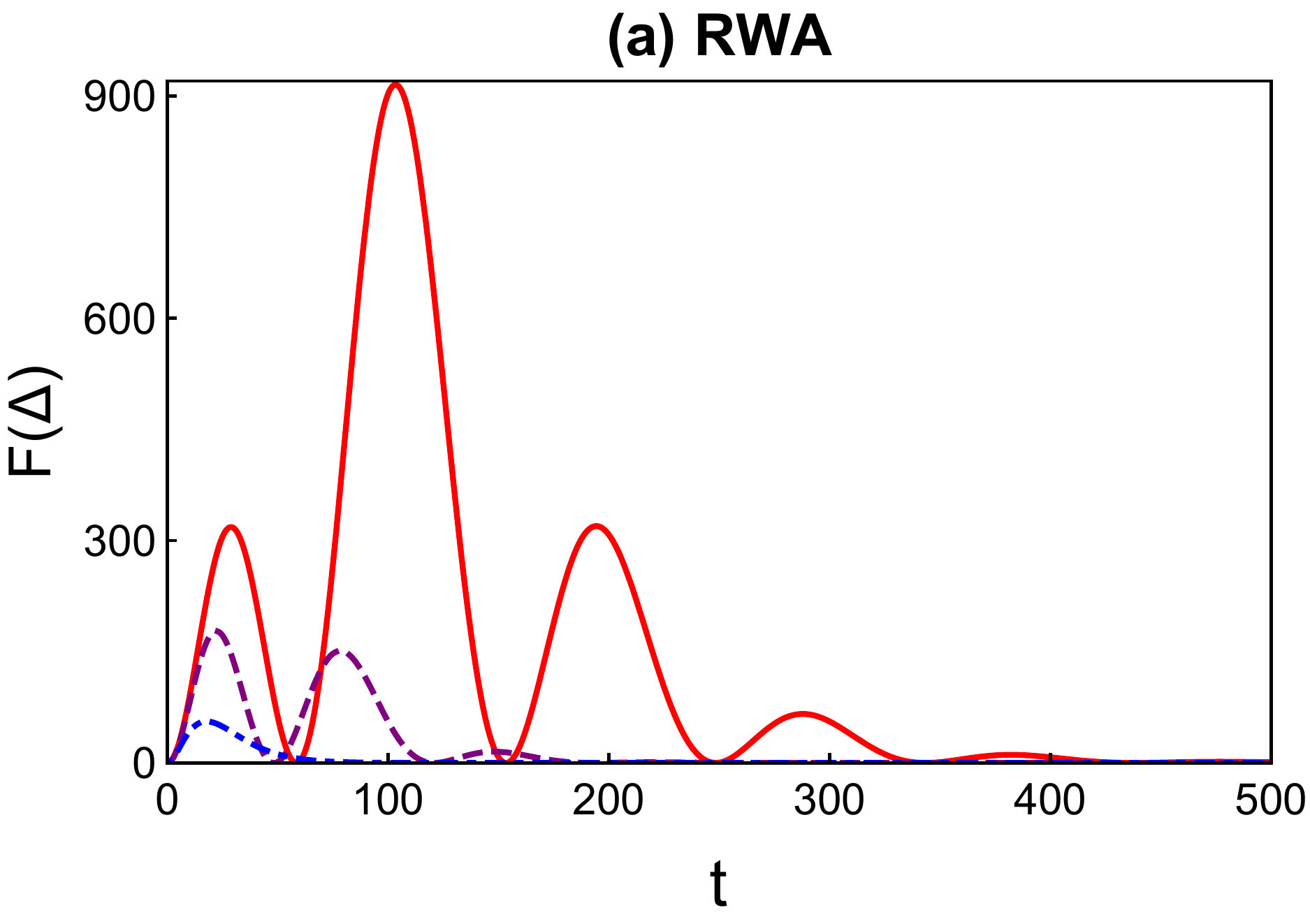}
\includegraphics[angle=0,width=5.5cm]{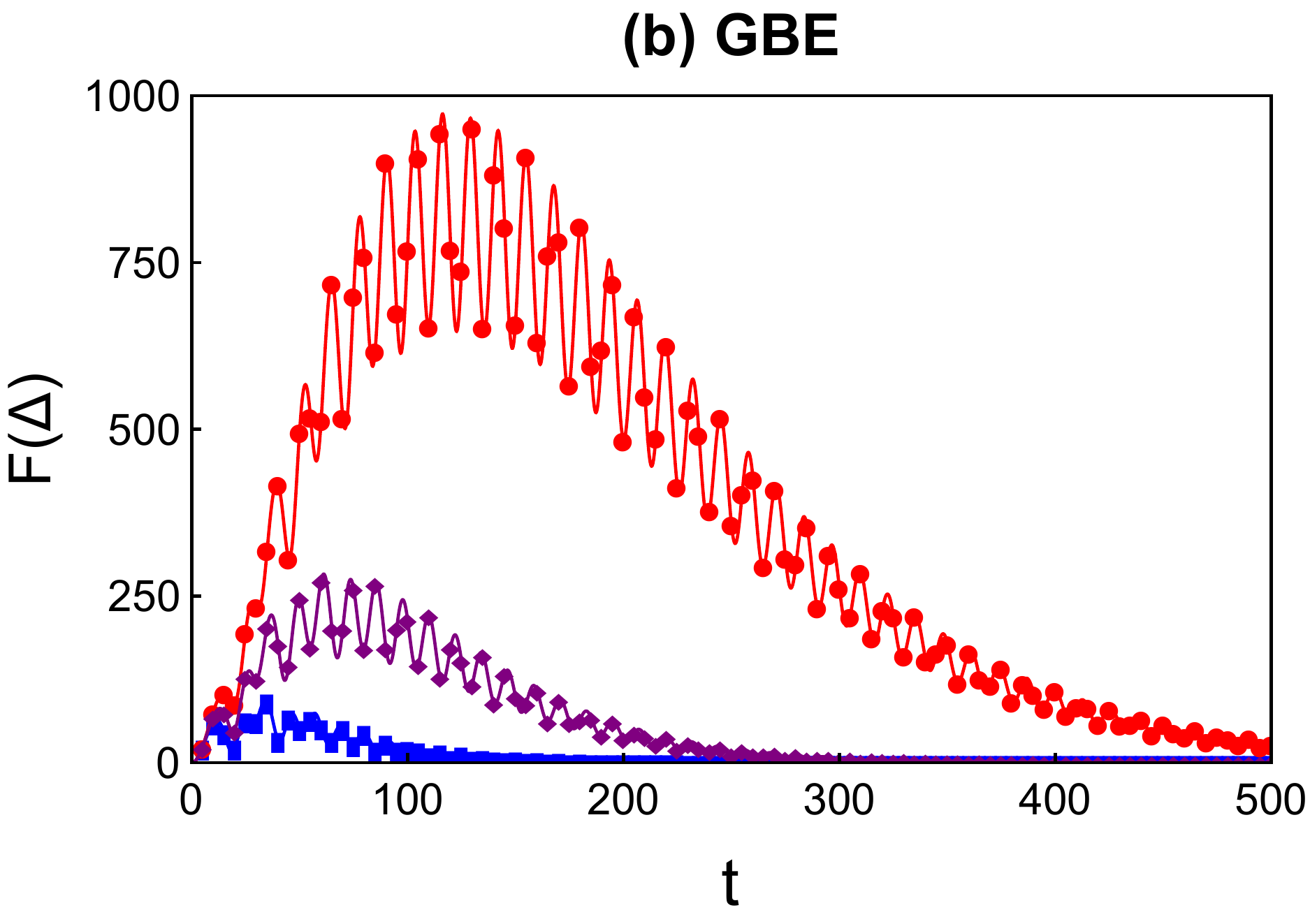}
\includegraphics[angle=0,width=5.5cm]{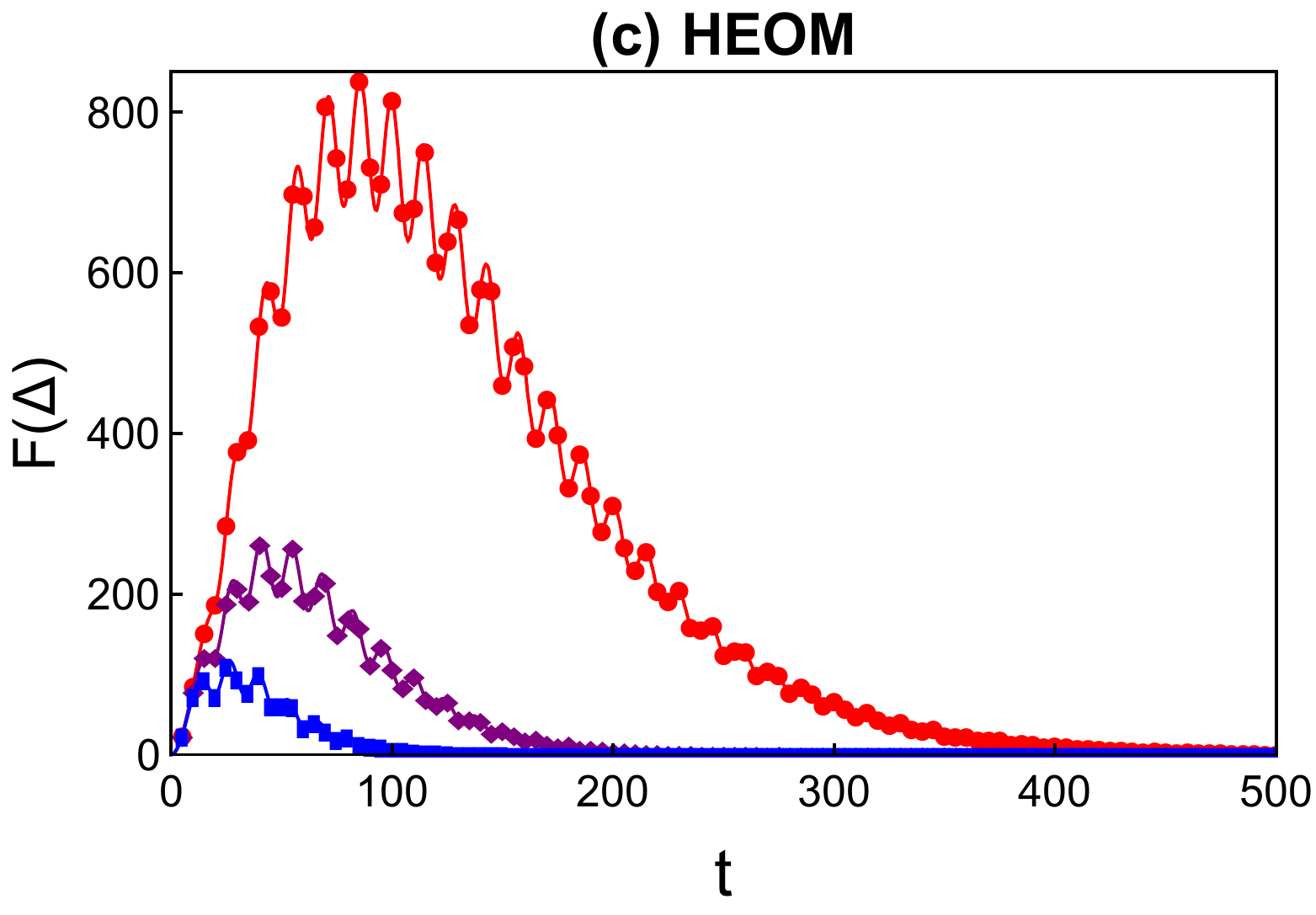}
\caption{(a) The QFI $F(\Delta)$ is plotted as a function of time in RWA case. Parameters are chosen as $\gamma=0.25\Gamma$ (red solid line), $\gamma=0.5\Gamma$ (purple dashed line) and $\gamma=5\Gamma$ (blue dotdashed line) with $\Gamma=0.1$ and $\Delta=1$. (b) The QFI $F(\Delta)$ from the GBE method. Parameters are chosen as $\gamma=0.25\Gamma$ (red circles), $\gamma=0.35\Gamma$ (purple rhombuses) and $\gamma=0.5\Gamma$ (blue rectangles) with $\Gamma=0.2$ and $\Delta=0.2$. (c) The QFI obtained by the numerical HEOM method. Parameters are chosen as $\gamma=0.25\Gamma$ (red circles), $\gamma=0.4\Gamma$ (purple rhombuses) and $\gamma=0.6\Gamma$ (blue rectangles) with $\Gamma=0.15$ and $\Delta=0.2$.}\label{fig:fig2}
\end{figure*}

\section{Results}\label{sec:sec4}

In this section, we study the influence of $\alpha(t)$ and $\hat{\mathcal{S}}$ on the estimation precision of $\Delta$ in a dissipative environment. During the numerical calculations to the exact QFI using the HEOM method, one needs to handle the first order derivative to the parameter $\Delta$, namely $\partial_{\Delta}\langle\hat{\sigma}_{i}(t)\rangle$ (see Eq.~(\ref{eq:eq3})). In this paper, the derivative for an arbitrary $\theta$-dependent function $f_{\theta}$ is numerically performed by adopting the following finite difference method
\begin{equation}\label{eq:eq18}
\frac{\partial f_{\theta}}{\partial\theta}\simeq\frac{-f_{\theta+2\epsilon}+8f_{\theta+\epsilon}-8f_{\theta-\epsilon}+f_{\theta-2\epsilon}}{12\epsilon}.
\end{equation}
In our numerical simulations, we set $\epsilon/\theta=10^{-5}$, which provides a very good accuracy for finite-difference approximations. In this section, we assume the initial state of the quantum probe is given by $\frac{1}{\sqrt{2}}(|+\rangle+|-\rangle)$.

\subsection{Effect of non-Markovainity}

We first study the environmental memory effect on the noisy estimation precision. As discussion in~\ref{subsec:subsec3d}, by manipulating the ratio of $\gamma/\Gamma$, the degree of non-Markovianity in the decoherence channel changes drastically. This feature is beneficial for us to explore the connection between the non-Markovianity and the estimation precision in a dissipative environment.

In the RWA case, one can derive a very simple expression of the QFI with respect to the parameter $\Delta$ as $F_{\mathrm{RWA}}(\Delta)=t^{2}\mathcal{G}_{t}^{2}$. With this expression at hand, it is very easy to check that the value of QFI can be boosted by decreasing the ratio of $\gamma/\Gamma$. This result implies the non-Markovianity may increase the estimation precision, which is in agreement with the results of Refs.~\cite{PhysRevLett.109.233601,PhysRevA.88.035806}. Moreover, in the non-Markovian regime, we observe that the QFI oscillates with time and exhibits a collapse-and-revival phenomenon before complete disappearance. The same result is also reported in Ref.~\cite{PhysRevA.88.035806}, and can be regarded as an evidence of reversed information flow from the environment back to the probe. Going beyond the RWA, the numerical performances from the GME and the HEOM tell us the same conclusion, see Fig.~\ref{fig:fig2} (b) and (c). Thus, one can conclude that the environmental non-Markovian effect can effectively improve the estimation precision regardless of whether the counter-rotating-wave terms are taken into account. In this sense, our result is a non-trivial generalization of Ref.~\cite{PhysRevA.88.035806} in which only the RWA case is considered.

\subsection{Dephasing versus relaxation}

\begin{figure}
\centering
\includegraphics[angle=0,width=4.25cm]{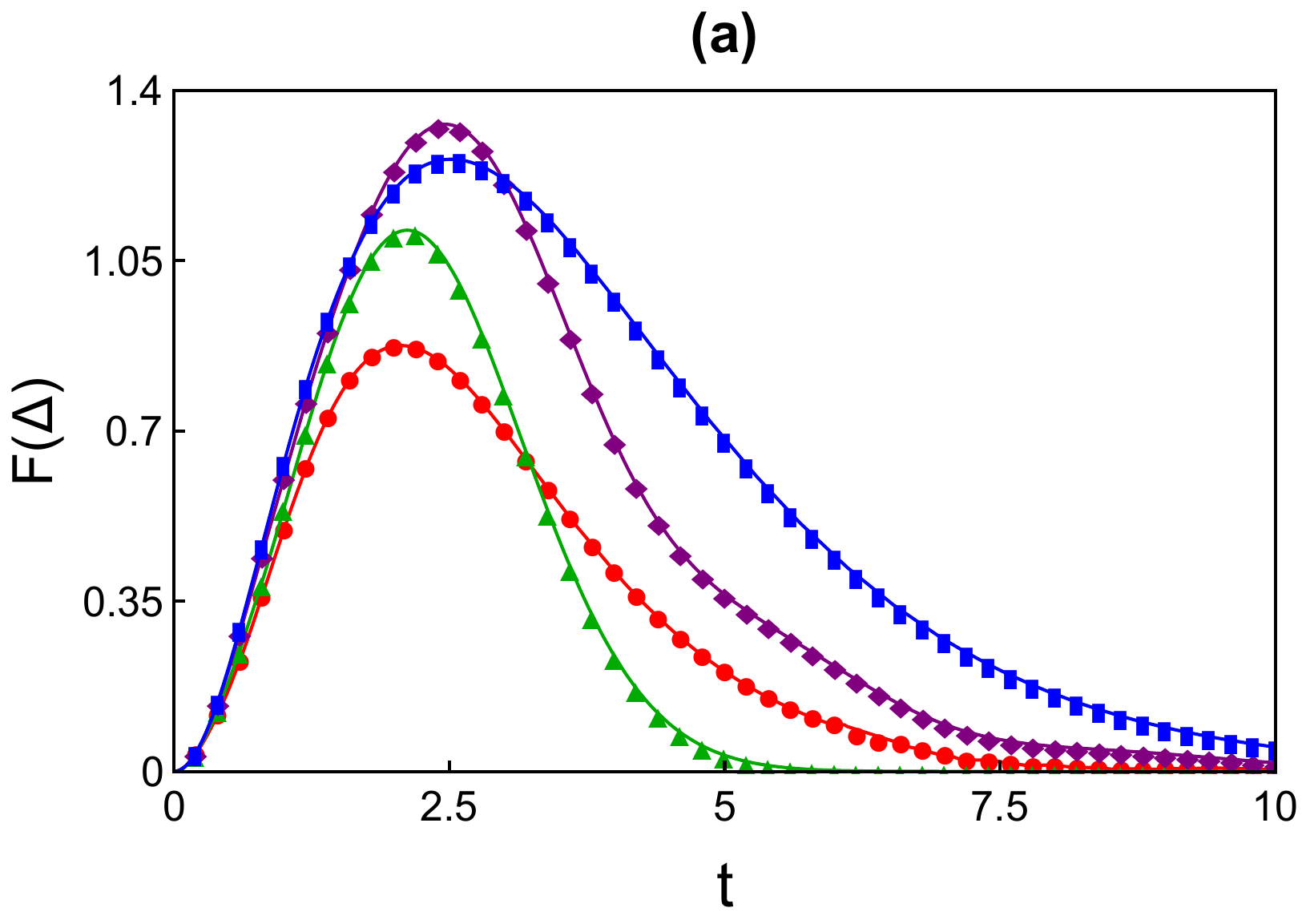}
\includegraphics[angle=0,width=4.25cm]{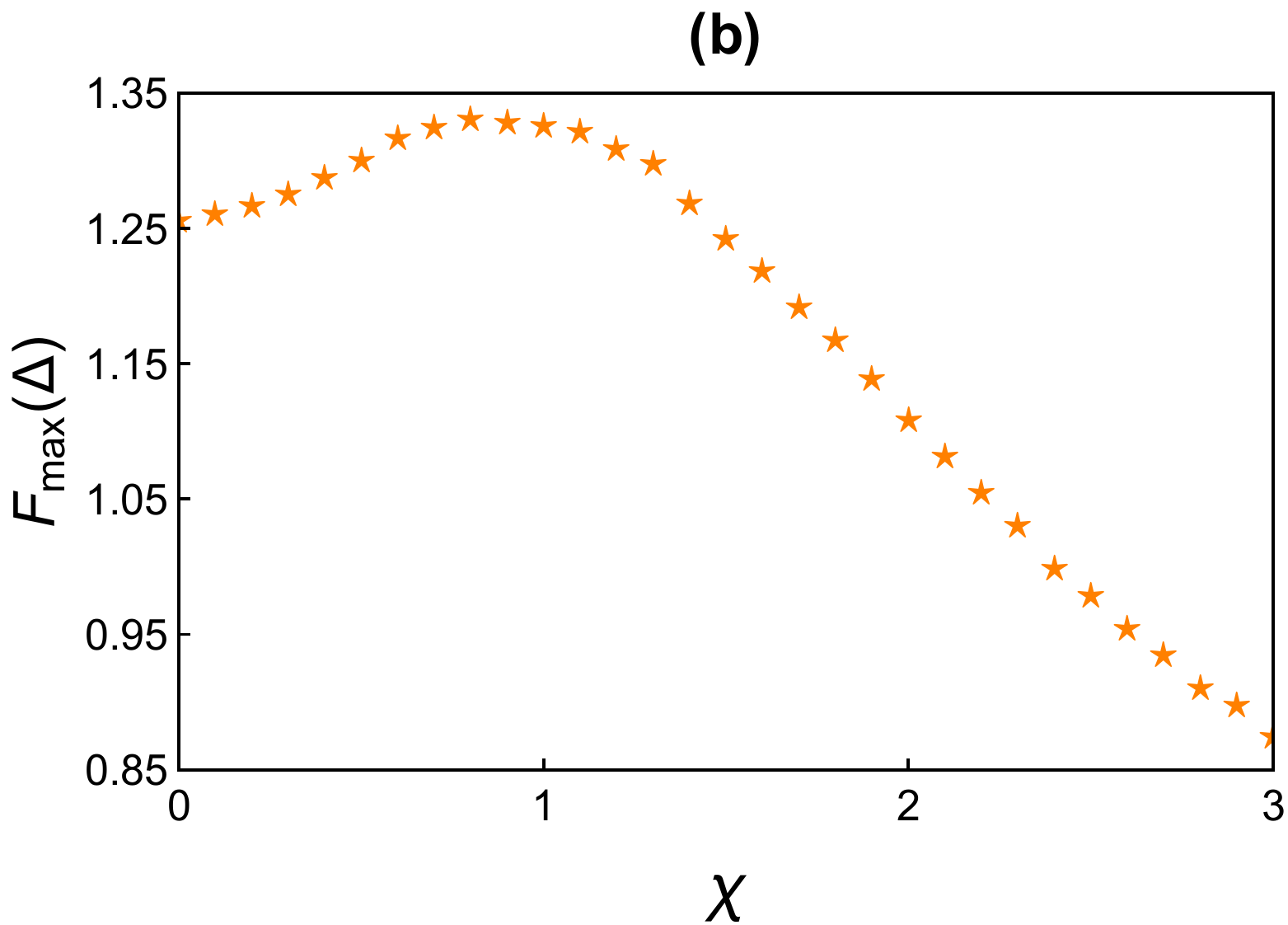}
\includegraphics[angle=0,width=4.25cm]{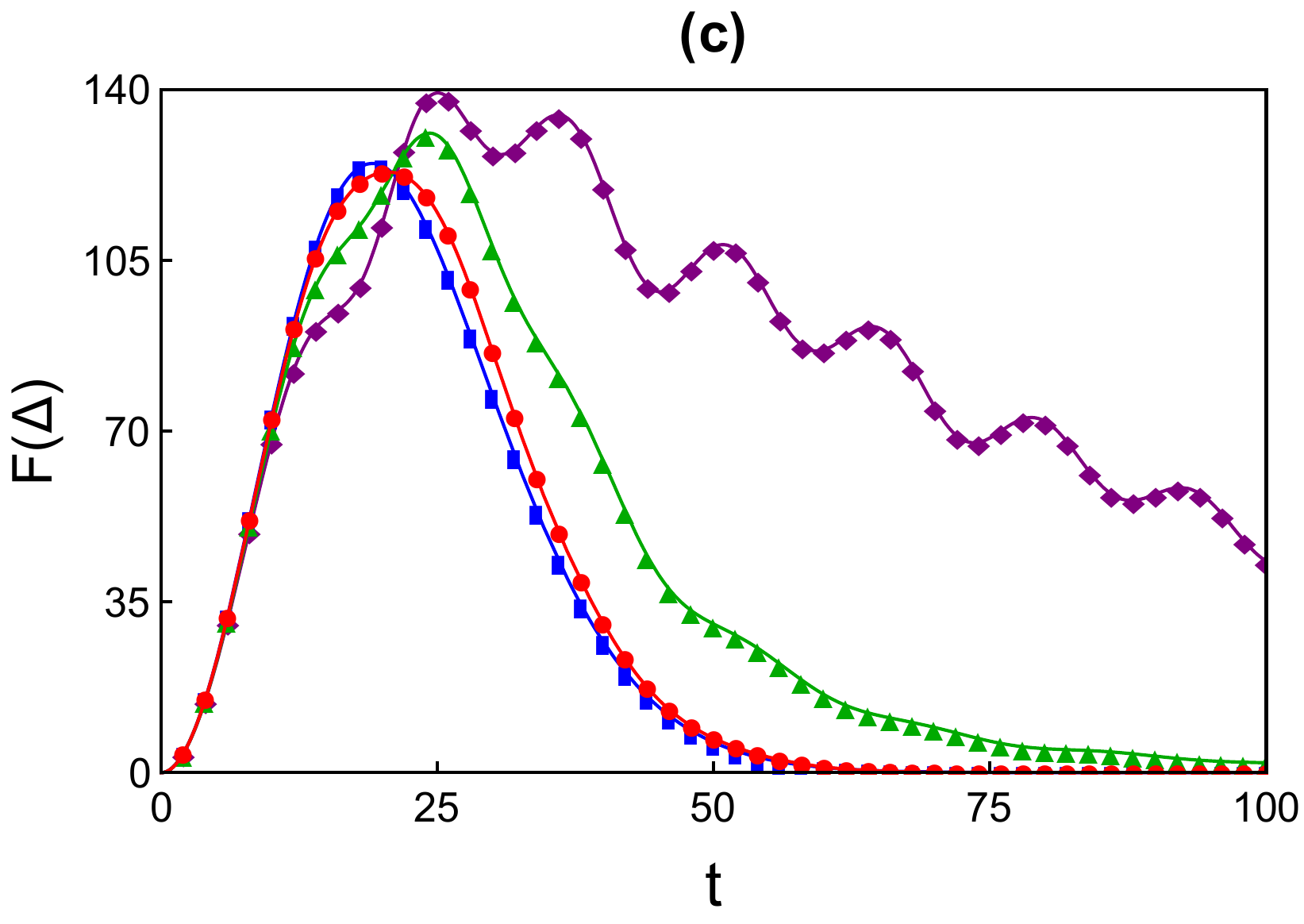}
\includegraphics[angle=0,width=4.25cm]{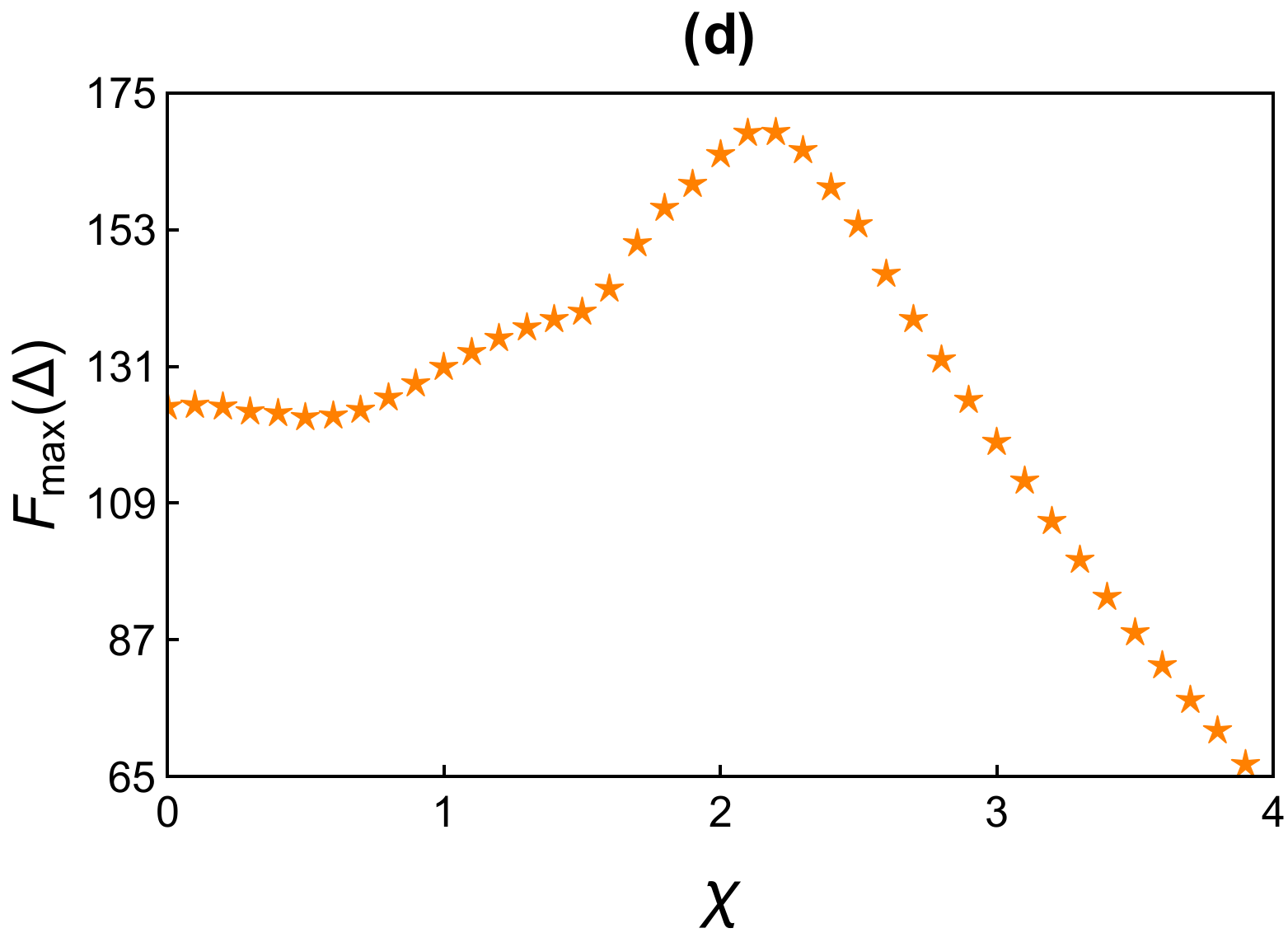}
\caption{(a) The QFI $F(\Delta)$ is plotted as a function of $t$ by making use of the HEOM method with different values of $\chi$: $\chi=0$ (blue rectangles), $\chi=0.75$ (purple rhombuses), $\chi=2$ (green triangles) and $\chi=3$ (red circles). (b) The maximum QFI with respect to time versus $\chi$. Parameters are chosen as $\gamma=10\Gamma$, $\Gamma=0.2\Delta$ and $\Delta=1$. (c) The QFI $F(\Delta)$ is displayed as a function of $t$ in the non-Markovian regime with different values of $\chi$: $\chi=0$ (blue rectangles), $\chi=1.5$ (purple rhombuses), $\chi=1$ (green triangles) and $\chi=0.5$ (red circles). (d) The maximum QFI versus $\chi$ in the non-Markovian regime. Parameters are chosen as $\gamma=0.3\Gamma$, $\Gamma=0.3\Delta$ and $\Delta=0.25$. }\label{fig:fig3}
\end{figure}

Generally speaking, the specific form of the probe-environment operator $\hat{\mathcal{S}}$ fully determines the decoherence channel. When $[\hat{\mathcal{S}},\hat{H}_{\mathrm{s}}]=0$, the probe suffers a pure dephasing decoherence mechanism and only off-diagonal elements of $\varrho_{\mathrm{s}}(t)$ decay during the time evolution. If $[\hat{\mathcal{S}},\hat{H}_{\mathrm{s}}]\neq 0$, the decoherence channel of the probe is relaxation, which results in the dissipation of qubit's energy. An interesting question arises here: what is the influence of the type of decoherence channel on the estimation performance? To address this problem, we generalize our discussion to a more general situation $\hat{\mathcal{S}}=\hat{\sigma}_{x}+\chi\hat{\sigma}_{z}$, where $\chi$ is a tunable real parameter~\cite{doi:10.1063/1.4950888}. Here, both parallel interaction case $\chi=0$ and perpendicular interaction case $\chi\neq0$ are included in the above expression of $\hat{\mathcal{S}}$, which can give rise to a much richer decoherence phenomenon. Such $\mathcal{\hat{S}}$ can be physically realized in atomic gas or quantum dot system, in which the atoms or electron spins are relaxed by their surrounding phonons (say, spontaneous emission process), meanwhile the dephasing process is generated by the random fluctuations of an external electromagnetic field~\cite{PhysRevLett.96.097009,doi:10.1063/1.5039891}.

Making use of the HEOM approach, which is independent of the specific form of operator $\mathcal{\hat{S}}$, we can numerically obtain the value of QFI. From Fig.~\ref{fig:fig3}, we find the influence of $\chi$ on the QFI is not evident in the short-time regime. However, as time increases, the effect of $\chi$ becomes no longer negligible. Maximizing the QFI over time, one can see the maximum QFI $F_{\max}(\Delta)$ is quite sensitive to the value of $\chi$: when $\chi$ is small, the introduction of the perpendicular interaction is favorable for obtaining a larger $F_{\max}(\Delta)$; after reaching a local maximum value, $F_{\max}(\Delta)$ gradually decreases as $\chi$ further increases. This result implies that the pure dephasing decoherence mechanism is \emph{not} the best choice for obtaining the maximum precision estimation, which is consistent with the result reported in Ref.~\cite{tamascelli2020quantum}. From Fig.~\ref{fig:fig3} (b) and (d), one can observe that $F_{\max}(\Delta)$ can be smaller than that of the pure dephasing case in the large-$\chi$ regime, which suggests there exists an optimal $\chi$ maximizing the value of QFI. Thus, we draw a conclusion that the performance of noisy parameter estimation can be enhanced by engineering the form of prob-environment coupling.

\section{Summary}\label{sec:sec5}

In summary, by employing the HEOM method, we have investigated the ultimate achievable limit to a qubit-probe's frequency estimation in a dissipative bosonic environment. Compared with two other approaches, it is found that the non-Markovian memory effect induced by the environment can remarkably boost the estimation precision, regardless of RWA or non-RWA cases. This is good news for a practical quantum sensing protocol, because the actual noisy environment is complicated and non-Markovian, compared with the over-simplified memoryless approximation used in certain theoretical treatments. We also reveal that the pure dephasing is not the optimal decoherence mechanism to obtain the maximum estimation precision. By introducing a perpendicular qubit-environment interaction, the estimation performance can be improved. Furthermore, by adjusting the value of $\chi$ to change the weight of the perpendicular interaction in the $\hat{\mathcal{S}}$ operator, one can attain a larger value of QFI. Due to the fact that both the specific forms of $\alpha(t)$ and $\mathcal{\hat{S}}$ play important roles in determining reduced dynamical behavior of the qubit-probe, our result implies the noisy parameter estimation precision can be optimized by controlling the decoherence mechanism.

Though these results are achieved in the Ornstein-Uhlenbeck auto-correlation function case, thanks to the rapid development of HEOM method, our analysis of noisy parameter estimation can be generalized to other auto-correlation functions. For example, as reported in Refs.~\cite{PhysRevA.98.012110,PhysRevA.98.032116,doi:10.1063/1.4936924,PhysRevB.95.214308}, the HEOM method has been extended to arbitrary spectral density function as well as finite temperature environment situation. Moreover, it has been reported that the HEOM method can be extended to simulate the dissipative dynamics of a few-level system embedded in a fermionic environment~\cite{PhysRevA.96.032125,Suess2015,doi:10.1063/1.5136093} or a spin environment~\cite{doi:10.1063/1.5018725,doi:10.1063/1.5018726}. It would be very interesting to extrapolate our study to these more general situations.

Finally, due to the comprehensive utilizations of the qubit-based quantum sensor, our study provides a means of designing an optimal estimation scheme to characterize a parameter of interest in a noisy environment. The strategy explored in this paper might have certain potential applications in the researches of quantum metrology and quantum sensing.

\section{Acknowledgments}

W. Wu wishes to thank Dr. S.-Y. Bai, Prof. H.-G. Luo and Prof. J.-H. An for many useful discussions. This work is supported by the National Natural Science Foundation (Grant No. 11704025).

\section{Appendix}

In this appendix, we would like to show how to derive Eq.~(\ref{eq:eq6}) from the common Schr$\ddot{\mathrm{o}}$dinger equation $\partial_{t}|\Psi_{\mathrm{sb}}(t)\rangle=-i\hat{H}|\Psi_{\mathrm{sb}}(t)\rangle$. The whole probe-environment Hamiltonian in the interaction picture with respect to the environment reads
\begin{equation}\label{eq:eq19}
\hat{H}(t)=\hat{H}_{\mathrm{s}}+\mathcal{\hat{S}}\sum_{k}\Big{(}g_{k}\hat{b}_{k}^{\dagger}e^{i\omega_{k}t}+g_{k}\hat{b}_{k}e^{-i\omega_{k}t}\Big{)}.
\end{equation}
Substituting $\hat{H}(t)$ into the standard Schr$\ddot{\mathrm{o}}$dinger equation, we have
\begin{equation}\label{eq:eq20}
\partial_{t}|\Psi_{\mathrm{sb}}(t)\rangle=-i\bigg{[}\hat{H}_{\mathrm{s}}+\mathcal{\hat{S}}\sum_{k}g_{k}\hat{b}_{k}^{\dagger}e^{i\omega_{k}t}+g_{k}\hat{b}_{k}e^{-i\omega_{k}t}\bigg{]}|\Psi_{\mathrm{sb}}(t)\rangle.
\end{equation}
Then, we employe the Bargmann coherent state $|\mathbf{z}\rangle=\bigotimes_{k}|z_{k}\rangle$ with $|z_{k}\rangle\equiv e^{z_{k}\hat{b}_{k}^{\dagger}}|0_{k}\rangle$ to reexpress Eq.~(\ref{eq:eq20}). By left-multiplying the Bargmann coherent state $\langle \mathbf{z}|$ on both sides of Eq.~(\ref{eq:eq20}), one can find
\begin{equation}\label{eq:eq21}
\begin{split}
\partial_{t}&\langle \mathbf{z}|\Psi_{\mathrm{sb}}(t)\rangle=-i\hat{H}_{\mathrm{s}}\langle \mathbf{z}|\Psi_{\mathrm{sb}}(t)\rangle\\
&-i\mathcal{\hat{S}}\langle \mathbf{z}|\bigg{[}\sum_{k}g_{k}\hat{b}_{k}^{\dagger}e^{i\omega_{k}t}+g_{k}\hat{b}_{k}e^{-i\omega_{k}t}\bigg{]}|\Psi_{\mathrm{sb}}(t)\rangle.
\end{split}
\end{equation}
Next, using the following properties of the Bargmann coherent state
\begin{equation}\label{eq:eq22}
\hat{b}_{k}|z_{k}\rangle=z_{k}|z_{k}\rangle,~~\hat{b}_{k}^{\dagger}|z_{k}\rangle=\frac{\partial}{\partial z_{k}}|z_{k}\rangle,
\end{equation}
Eq.~(\ref{eq:eq21}) can be simplifies to
\begin{equation*}\label{eq:eq23}
\begin{split}
\partial_{t}|\psi_{t}(\mathbf{z}^{*})\rangle=\bigg{[}-i\hat{H}_{\mathrm{s}}+\mathcal{\hat{S}}\mathbf{z}_{t}^{*}-i\mathcal{\hat{S}}\sum_{k}g_{k}e^{-i\omega_{k}t}\frac{\partial}{\partial z_{k}^{*}}\bigg{]}|\psi_{t}(\mathbf{z}^{*})\rangle,
\end{split}
\end{equation*}
where $\textbf{z}_{t}\equiv i\sum_{k}g_{k}z_{k}e^{-i\omega_{k}t}$. The term $\frac{\partial}{\partial z_{k}^{*}}|\psi_{t}(\mathbf{z}^{*})\rangle$ can be cast as a functional derivative by making use of the functional chain rule~\cite{DIOSI1997569,STRUNZ2001237}
\begin{equation}\label{eq:eq24}
\begin{split}
\frac{\partial}{\partial z_{k}^{*}}|\psi_{t}(\mathbf{z}^{*})\rangle=\int_{0}^{t}d\tau\frac{\partial \mathbf{z}_{\tau}^{*}}{\partial z_{k}^{*}}\frac{\delta}{\delta \mathbf{z}_{\tau}^{*}}|\psi_{t}(\mathbf{z}^{*})\rangle.
\end{split}
\end{equation}
Finally, we have
\begin{equation*}\label{eq:eq25}
\begin{split}
\frac{\partial}{\partial t}|\psi_{t}(\textbf{z}^{*})\rangle=&-i\hat{H}_{\mathrm{s}}|\psi_{t}(\textbf{z}^{*})\rangle+\hat{\mathcal{S}}\textbf{z}_{t}^{*}|\psi_{t}(\textbf{z}^{*})\rangle\\
&-\hat{\mathcal{S}}\int_{0}^{t}d\tau\sum_{k}g_{k}^{2}e^{-i\omega_{k}(t-\tau)}\frac{\delta}{\delta \textbf{z}_{\tau}^{*}}|\psi_{t}(\textbf{z}^{*})\rangle,
\end{split}
\end{equation*}
which reproduces Eq.~(\ref{eq:eq6}) in the main text. Therefore, by defining the stochastic process $\mathbf{z}_{t}$ which originates from environmental degrees of freedom, the standard Schr$\ddot{\mathrm{o}}$dinger equation can be converted into the stochastic quantum state diffusion equation.

\bibliography{reference}

\begin{thebibliography}{83}%
\makeatletter
\providecommand \@ifxundefined [1]{%
 \@ifx{#1\undefined}
}%
\providecommand \@ifnum [1]{%
 \ifnum #1\expandafter \@firstoftwo
 \else \expandafter \@secondoftwo
 \fi
}%
\providecommand \@ifx [1]{%
 \ifx #1\expandafter \@firstoftwo
 \else \expandafter \@secondoftwo
 \fi
}%
\providecommand \natexlab [1]{#1}%
\providecommand \enquote  [1]{``#1''}%
\providecommand \bibnamefont  [1]{#1}%
\providecommand \bibfnamefont [1]{#1}%
\providecommand \citenamefont [1]{#1}%
\providecommand \href@noop [0]{\@secondoftwo}%
\providecommand \href [0]{\begingroup \@sanitize@url \@href}%
\providecommand \@href[1]{\@@startlink{#1}\@@href}%
\providecommand \@@href[1]{\endgroup#1\@@endlink}%
\providecommand \@sanitize@url [0]{\catcode `\\12\catcode `\$12\catcode
  `\&12\catcode `\#12\catcode `\^12\catcode `\_12\catcode `\%12\relax}%
\providecommand \@@startlink[1]{}%
\providecommand \@@endlink[0]{}%
\providecommand \url  [0]{\begingroup\@sanitize@url \@url }%
\providecommand \@url [1]{\endgroup\@href {#1}{\urlprefix }}%
\providecommand \urlprefix  [0]{URL }%
\providecommand \Eprint [0]{\href }%
\providecommand \doibase [0]{http://dx.doi.org/}%
\providecommand \selectlanguage [0]{\@gobble}%
\providecommand \bibinfo  [0]{\@secondoftwo}%
\providecommand \bibfield  [0]{\@secondoftwo}%
\providecommand \translation [1]{[#1]}%
\providecommand \BibitemOpen [0]{}%
\providecommand \bibitemStop [0]{}%
\providecommand \bibitemNoStop [0]{.\EOS\space}%
\providecommand \EOS [0]{\spacefactor3000\relax}%
\providecommand \BibitemShut  [1]{\csname bibitem#1\endcsname}%
\let\auto@bib@innerbib\@empty
\bibitem [{\citenamefont {Tse}\ \emph {et~al.}(2019)\citenamefont {Tse} \emph
  {et~al.}}]{PhysRevLett.123.231107}%
  \BibitemOpen
  \bibfield  {author} {\bibinfo {author} {\bibfnamefont {M.}~\bibnamefont
  {Tse}} \emph {et~al.},\ }\bibfield  {title} {\enquote {\bibinfo {title}
  {Quantum-enhanced advanced ligo detectors in the era of gravitational-wave
  astronomy},}\ }\href {\doibase 10.1103/PhysRevLett.123.231107} {\bibfield
  {journal} {\bibinfo  {journal} {Phys. Rev. Lett.}\ }\textbf {\bibinfo
  {volume} {123}},\ \bibinfo {pages} {231107} (\bibinfo {year}
  {2019})}\BibitemShut {NoStop}%
\bibitem [{\citenamefont {Acernese}\ \emph {et~al.}(2019)\citenamefont
  {Acernese} \emph {et~al.}}]{PhysRevLett.123.231108}%
  \BibitemOpen
  \bibfield  {author} {\bibinfo {author} {\bibfnamefont {F.}~\bibnamefont
  {Acernese}} \emph {et~al.},\ }\bibfield  {title} {\enquote {\bibinfo {title}
  {Increasing the astrophysical reach of the advanced virgo detector via the
  application of squeezed vacuum states of light},}\ }\href {\doibase
  10.1103/PhysRevLett.123.231108} {\bibfield  {journal} {\bibinfo  {journal}
  {Phys. Rev. Lett.}\ }\textbf {\bibinfo {volume} {123}},\ \bibinfo {pages}
  {231108} (\bibinfo {year} {2019})}\BibitemShut {NoStop}%
\bibitem [{\citenamefont {Xu}\ and\ \citenamefont
  {Holland}(2015)}]{PhysRevLett.114.103601}%
  \BibitemOpen
  \bibfield  {author} {\bibinfo {author} {\bibfnamefont {Minghui}\ \bibnamefont
  {Xu}}\ and\ \bibinfo {author} {\bibfnamefont {M.~J.}\ \bibnamefont
  {Holland}},\ }\bibfield  {title} {\enquote {\bibinfo {title} {Conditional
  ramsey spectroscopy with synchronized atoms},}\ }\href {\doibase
  10.1103/PhysRevLett.114.103601} {\bibfield  {journal} {\bibinfo  {journal}
  {Phys. Rev. Lett.}\ }\textbf {\bibinfo {volume} {114}},\ \bibinfo {pages}
  {103601} (\bibinfo {year} {2015})}\BibitemShut {NoStop}%
\bibitem [{\citenamefont {Xu}\ \emph {et~al.}(2014)\citenamefont {Xu},
  \citenamefont {Tieri}, \citenamefont {Fine}, \citenamefont {Thompson},\ and\
  \citenamefont {Holland}}]{PhysRevLett.113.154101}%
  \BibitemOpen
  \bibfield  {author} {\bibinfo {author} {\bibfnamefont {Minghui}\ \bibnamefont
  {Xu}}, \bibinfo {author} {\bibfnamefont {D.~A.}\ \bibnamefont {Tieri}},
  \bibinfo {author} {\bibfnamefont {E.~C.}\ \bibnamefont {Fine}}, \bibinfo
  {author} {\bibfnamefont {James~K.}\ \bibnamefont {Thompson}}, \ and\ \bibinfo
  {author} {\bibfnamefont {M.~J.}\ \bibnamefont {Holland}},\ }\bibfield
  {title} {\enquote {\bibinfo {title} {Synchronization of two ensembles of
  atoms},}\ }\href {\doibase 10.1103/PhysRevLett.113.154101} {\bibfield
  {journal} {\bibinfo  {journal} {Phys. Rev. Lett.}\ }\textbf {\bibinfo
  {volume} {113}},\ \bibinfo {pages} {154101} (\bibinfo {year}
  {2014})}\BibitemShut {NoStop}%
\bibitem [{\citenamefont {Correa}\ \emph {et~al.}(2015)\citenamefont {Correa},
  \citenamefont {Mehboudi}, \citenamefont {Adesso},\ and\ \citenamefont
  {Sanpera}}]{PhysRevLett.114.220405}%
  \BibitemOpen
  \bibfield  {author} {\bibinfo {author} {\bibfnamefont {Luis~A.}\ \bibnamefont
  {Correa}}, \bibinfo {author} {\bibfnamefont {Mohammad}\ \bibnamefont
  {Mehboudi}}, \bibinfo {author} {\bibfnamefont {Gerardo}\ \bibnamefont
  {Adesso}}, \ and\ \bibinfo {author} {\bibfnamefont {Anna}\ \bibnamefont
  {Sanpera}},\ }\bibfield  {title} {\enquote {\bibinfo {title} {Individual
  quantum probes for optimal thermometry},}\ }\href {\doibase
  10.1103/PhysRevLett.114.220405} {\bibfield  {journal} {\bibinfo  {journal}
  {Phys. Rev. Lett.}\ }\textbf {\bibinfo {volume} {114}},\ \bibinfo {pages}
  {220405} (\bibinfo {year} {2015})}\BibitemShut {NoStop}%
\bibitem [{\citenamefont {Hovhannisyan}\ and\ \citenamefont
  {Correa}(2018)}]{PhysRevB.98.045101}%
  \BibitemOpen
  \bibfield  {author} {\bibinfo {author} {\bibfnamefont {Karen~V.}\
  \bibnamefont {Hovhannisyan}}\ and\ \bibinfo {author} {\bibfnamefont
  {Luis~A.}\ \bibnamefont {Correa}},\ }\bibfield  {title} {\enquote {\bibinfo
  {title} {Measuring the temperature of cold many-body quantum systems},}\
  }\href {\doibase 10.1103/PhysRevB.98.045101} {\bibfield  {journal} {\bibinfo
  {journal} {Phys. Rev. B}\ }\textbf {\bibinfo {volume} {98}},\ \bibinfo
  {pages} {045101} (\bibinfo {year} {2018})}\BibitemShut {NoStop}%
\bibitem [{\citenamefont {Bouton}\ \emph {et~al.}(2020)\citenamefont {Bouton},
  \citenamefont {Nettersheim}, \citenamefont {Adam}, \citenamefont {Schmidt},
  \citenamefont {Mayer}, \citenamefont {Lausch}, \citenamefont {Tiemann},\ and\
  \citenamefont {Widera}}]{PhysRevX.10.011018}%
  \BibitemOpen
  \bibfield  {author} {\bibinfo {author} {\bibfnamefont {Quentin}\ \bibnamefont
  {Bouton}}, \bibinfo {author} {\bibfnamefont {Jens}\ \bibnamefont
  {Nettersheim}}, \bibinfo {author} {\bibfnamefont {Daniel}\ \bibnamefont
  {Adam}}, \bibinfo {author} {\bibfnamefont {Felix}\ \bibnamefont {Schmidt}},
  \bibinfo {author} {\bibfnamefont {Daniel}\ \bibnamefont {Mayer}}, \bibinfo
  {author} {\bibfnamefont {Tobias}\ \bibnamefont {Lausch}}, \bibinfo {author}
  {\bibfnamefont {Eberhard}\ \bibnamefont {Tiemann}}, \ and\ \bibinfo {author}
  {\bibfnamefont {Artur}\ \bibnamefont {Widera}},\ }\bibfield  {title}
  {\enquote {\bibinfo {title} {Single-atom quantum probes for ultracold gases
  boosted by nonequilibrium spin dynamics},}\ }\href {\doibase
  10.1103/PhysRevX.10.011018} {\bibfield  {journal} {\bibinfo  {journal} {Phys.
  Rev. X}\ }\textbf {\bibinfo {volume} {10}},\ \bibinfo {pages} {011018}
  (\bibinfo {year} {2020})}\BibitemShut {NoStop}%
\bibitem [{\citenamefont {Herrera-Mart\'{\i}}\ \emph
  {et~al.}(2015)\citenamefont {Herrera-Mart\'{\i}}, \citenamefont {Gefen},
  \citenamefont {Aharonov}, \citenamefont {Katz},\ and\ \citenamefont
  {Retzker}}]{PhysRevLett.115.200501}%
  \BibitemOpen
  \bibfield  {author} {\bibinfo {author} {\bibfnamefont {David~A.}\
  \bibnamefont {Herrera-Mart\'{\i}}}, \bibinfo {author} {\bibfnamefont {Tuvia}\
  \bibnamefont {Gefen}}, \bibinfo {author} {\bibfnamefont {Dorit}\ \bibnamefont
  {Aharonov}}, \bibinfo {author} {\bibfnamefont {Nadav}\ \bibnamefont {Katz}},
  \ and\ \bibinfo {author} {\bibfnamefont {Alex}\ \bibnamefont {Retzker}},\
  }\bibfield  {title} {\enquote {\bibinfo {title} {Quantum
  error-correction-enhanced magnetometer overcoming the limit imposed by
  relaxation},}\ }\href {\doibase 10.1103/PhysRevLett.115.200501} {\bibfield
  {journal} {\bibinfo  {journal} {Phys. Rev. Lett.}\ }\textbf {\bibinfo
  {volume} {115}},\ \bibinfo {pages} {200501} (\bibinfo {year}
  {2015})}\BibitemShut {NoStop}%
\bibitem [{\citenamefont {Baumgart}\ \emph {et~al.}(2016)\citenamefont
  {Baumgart}, \citenamefont {Cai}, \citenamefont {Retzker}, \citenamefont
  {Plenio},\ and\ \citenamefont {Wunderlich}}]{PhysRevLett.116.240801}%
  \BibitemOpen
  \bibfield  {author} {\bibinfo {author} {\bibfnamefont {I.}~\bibnamefont
  {Baumgart}}, \bibinfo {author} {\bibfnamefont {J.-M.}\ \bibnamefont {Cai}},
  \bibinfo {author} {\bibfnamefont {A.}~\bibnamefont {Retzker}}, \bibinfo
  {author} {\bibfnamefont {M.~B.}\ \bibnamefont {Plenio}}, \ and\ \bibinfo
  {author} {\bibfnamefont {Ch.}\ \bibnamefont {Wunderlich}},\ }\bibfield
  {title} {\enquote {\bibinfo {title} {Ultrasensitive magnetometer using a
  single atom},}\ }\href {\doibase 10.1103/PhysRevLett.116.240801} {\bibfield
  {journal} {\bibinfo  {journal} {Phys. Rev. Lett.}\ }\textbf {\bibinfo
  {volume} {116}},\ \bibinfo {pages} {240801} (\bibinfo {year}
  {2016})}\BibitemShut {NoStop}%
\bibitem [{\citenamefont {Bhattacharjee}\ \emph {et~al.}(2020)\citenamefont
  {Bhattacharjee}, \citenamefont {Bhattacharya}, \citenamefont {Niedenzu},
  \citenamefont {Mukherjee},\ and\ \citenamefont {Dutta}}]{Bhattacharjee_2020}%
  \BibitemOpen
  \bibfield  {author} {\bibinfo {author} {\bibfnamefont {Sourav}\ \bibnamefont
  {Bhattacharjee}}, \bibinfo {author} {\bibfnamefont {Utso}\ \bibnamefont
  {Bhattacharya}}, \bibinfo {author} {\bibfnamefont {Wolfgang}\ \bibnamefont
  {Niedenzu}}, \bibinfo {author} {\bibfnamefont {Victor}\ \bibnamefont
  {Mukherjee}}, \ and\ \bibinfo {author} {\bibfnamefont {Amit}\ \bibnamefont
  {Dutta}},\ }\bibfield  {title} {\enquote {\bibinfo {title} {Quantum
  magnetometry using two-stroke thermal machines},}\ }\href {\doibase
  10.1088/1367-2630/ab61d6} {\bibfield  {journal} {\bibinfo  {journal} {New
  Journal of Physics}\ }\textbf {\bibinfo {volume} {22}},\ \bibinfo {pages}
  {013024} (\bibinfo {year} {2020})}\BibitemShut {NoStop}%
\bibitem [{\citenamefont {Nagata}\ \emph {et~al.}(2007)\citenamefont {Nagata},
  \citenamefont {Okamoto}, \citenamefont {O{\textquoteright}Brien},
  \citenamefont {Sasaki},\ and\ \citenamefont {Takeuchi}}]{Nagata726}%
  \BibitemOpen
  \bibfield  {author} {\bibinfo {author} {\bibfnamefont {Tomohisa}\
  \bibnamefont {Nagata}}, \bibinfo {author} {\bibfnamefont {Ryo}\ \bibnamefont
  {Okamoto}}, \bibinfo {author} {\bibfnamefont {Jeremy~L.}\ \bibnamefont
  {O{\textquoteright}Brien}}, \bibinfo {author} {\bibfnamefont {Keiji}\
  \bibnamefont {Sasaki}}, \ and\ \bibinfo {author} {\bibfnamefont {Shigeki}\
  \bibnamefont {Takeuchi}},\ }\bibfield  {title} {\enquote {\bibinfo {title}
  {Beating the standard quantum limit with four-entangled photons},}\ }\href
  {\doibase 10.1126/science.1138007} {\bibfield  {journal} {\bibinfo  {journal}
  {Science}\ }\textbf {\bibinfo {volume} {316}},\ \bibinfo {pages} {726--729}
  (\bibinfo {year} {2007})}\BibitemShut {NoStop}%
\bibitem [{\citenamefont {Zou}\ \emph {et~al.}(2018)\citenamefont {Zou},
  \citenamefont {Wu}, \citenamefont {Liu}, \citenamefont {Luo}, \citenamefont
  {Guo}, \citenamefont {Cao}, \citenamefont {Tey},\ and\ \citenamefont
  {You}}]{Zou6381}%
  \BibitemOpen
  \bibfield  {author} {\bibinfo {author} {\bibfnamefont {Yi-Quan}\ \bibnamefont
  {Zou}}, \bibinfo {author} {\bibfnamefont {Ling-Na}\ \bibnamefont {Wu}},
  \bibinfo {author} {\bibfnamefont {Qi}~\bibnamefont {Liu}}, \bibinfo {author}
  {\bibfnamefont {Xin-Yu}\ \bibnamefont {Luo}}, \bibinfo {author}
  {\bibfnamefont {Shuai-Feng}\ \bibnamefont {Guo}}, \bibinfo {author}
  {\bibfnamefont {Jia-Hao}\ \bibnamefont {Cao}}, \bibinfo {author}
  {\bibfnamefont {Meng~Khoon}\ \bibnamefont {Tey}}, \ and\ \bibinfo {author}
  {\bibfnamefont {Li}~\bibnamefont {You}},\ }\bibfield  {title} {\enquote
  {\bibinfo {title} {Beating the classical precision limit with spin-1 dicke
  states of more than 10,000 atoms},}\ }\href {\doibase
  10.1073/pnas.1715105115} {\bibfield  {journal} {\bibinfo  {journal}
  {Proceedings of the National Academy of Sciences}\ }\textbf {\bibinfo
  {volume} {115}},\ \bibinfo {pages} {6381--6385} (\bibinfo {year}
  {2018})}\BibitemShut {NoStop}%
\bibitem [{\citenamefont {Zhang}\ \emph {et~al.}(2018)\citenamefont {Zhang},
  \citenamefont {Um}, \citenamefont {Lv}, \citenamefont {Zhang}, \citenamefont
  {Duan},\ and\ \citenamefont {Kim}}]{PhysRevLett.121.160502}%
  \BibitemOpen
  \bibfield  {author} {\bibinfo {author} {\bibfnamefont {Junhua}\ \bibnamefont
  {Zhang}}, \bibinfo {author} {\bibfnamefont {Mark}\ \bibnamefont {Um}},
  \bibinfo {author} {\bibfnamefont {Dingshun}\ \bibnamefont {Lv}}, \bibinfo
  {author} {\bibfnamefont {Jing-Ning}\ \bibnamefont {Zhang}}, \bibinfo {author}
  {\bibfnamefont {Lu-Ming}\ \bibnamefont {Duan}}, \ and\ \bibinfo {author}
  {\bibfnamefont {Kihwan}\ \bibnamefont {Kim}},\ }\bibfield  {title} {\enquote
  {\bibinfo {title} {Noon states of nine quantized vibrations in two radial
  modes of a trapped ion},}\ }\href {\doibase 10.1103/PhysRevLett.121.160502}
  {\bibfield  {journal} {\bibinfo  {journal} {Phys. Rev. Lett.}\ }\textbf
  {\bibinfo {volume} {121}},\ \bibinfo {pages} {160502} (\bibinfo {year}
  {2018})}\BibitemShut {NoStop}%
\bibitem [{\citenamefont {Haine}\ and\ \citenamefont
  {Szigeti}(2015)}]{PhysRevA.92.032317}%
  \BibitemOpen
  \bibfield  {author} {\bibinfo {author} {\bibfnamefont {Simon~A.}\
  \bibnamefont {Haine}}\ and\ \bibinfo {author} {\bibfnamefont {Stuart~S.}\
  \bibnamefont {Szigeti}},\ }\bibfield  {title} {\enquote {\bibinfo {title}
  {Quantum metrology with mixed states: When recovering lost information is
  better than never losing it},}\ }\href {\doibase 10.1103/PhysRevA.92.032317}
  {\bibfield  {journal} {\bibinfo  {journal} {Phys. Rev. A}\ }\textbf {\bibinfo
  {volume} {92}},\ \bibinfo {pages} {032317} (\bibinfo {year}
  {2015})}\BibitemShut {NoStop}%
\bibitem [{\citenamefont {Caves}(1981)}]{PhysRevD.23.1693}%
  \BibitemOpen
  \bibfield  {author} {\bibinfo {author} {\bibfnamefont {Carlton~M.}\
  \bibnamefont {Caves}},\ }\bibfield  {title} {\enquote {\bibinfo {title}
  {Quantum-mechanical noise in an interferometer},}\ }\href {\doibase
  10.1103/PhysRevD.23.1693} {\bibfield  {journal} {\bibinfo  {journal} {Phys.
  Rev. D}\ }\textbf {\bibinfo {volume} {23}},\ \bibinfo {pages} {1693--1708}
  (\bibinfo {year} {1981})}\BibitemShut {NoStop}%
\bibitem [{\citenamefont {Nolan}\ \emph {et~al.}(2017)\citenamefont {Nolan},
  \citenamefont {Szigeti},\ and\ \citenamefont
  {Haine}}]{PhysRevLett.119.193601}%
  \BibitemOpen
  \bibfield  {author} {\bibinfo {author} {\bibfnamefont {Samuel~P.}\
  \bibnamefont {Nolan}}, \bibinfo {author} {\bibfnamefont {Stuart~S.}\
  \bibnamefont {Szigeti}}, \ and\ \bibinfo {author} {\bibfnamefont {Simon~A.}\
  \bibnamefont {Haine}},\ }\bibfield  {title} {\enquote {\bibinfo {title}
  {Optimal and robust quantum metrology using interaction-based readouts},}\
  }\href {\doibase 10.1103/PhysRevLett.119.193601} {\bibfield  {journal}
  {\bibinfo  {journal} {Phys. Rev. Lett.}\ }\textbf {\bibinfo {volume} {119}},\
  \bibinfo {pages} {193601} (\bibinfo {year} {2017})}\BibitemShut {NoStop}%
\bibitem [{\citenamefont {Degen}\ \emph {et~al.}(2017)\citenamefont {Degen},
  \citenamefont {Reinhard},\ and\ \citenamefont
  {Cappellaro}}]{RevModPhys.89.035002}%
  \BibitemOpen
  \bibfield  {author} {\bibinfo {author} {\bibfnamefont {C.~L.}\ \bibnamefont
  {Degen}}, \bibinfo {author} {\bibfnamefont {F.}~\bibnamefont {Reinhard}}, \
  and\ \bibinfo {author} {\bibfnamefont {P.}~\bibnamefont {Cappellaro}},\
  }\bibfield  {title} {\enquote {\bibinfo {title} {Quantum sensing},}\ }\href
  {\doibase 10.1103/RevModPhys.89.035002} {\bibfield  {journal} {\bibinfo
  {journal} {Rev. Mod. Phys.}\ }\textbf {\bibinfo {volume} {89}},\ \bibinfo
  {pages} {035002} (\bibinfo {year} {2017})}\BibitemShut {NoStop}%
\bibitem [{\citenamefont {Pezz\`e}\ \emph {et~al.}(2018)\citenamefont
  {Pezz\`e}, \citenamefont {Smerzi}, \citenamefont {Oberthaler}, \citenamefont
  {Schmied},\ and\ \citenamefont {Treutlein}}]{RevModPhys.90.035005}%
  \BibitemOpen
  \bibfield  {author} {\bibinfo {author} {\bibfnamefont {Luca}\ \bibnamefont
  {Pezz\`e}}, \bibinfo {author} {\bibfnamefont {Augusto}\ \bibnamefont
  {Smerzi}}, \bibinfo {author} {\bibfnamefont {Markus~K.}\ \bibnamefont
  {Oberthaler}}, \bibinfo {author} {\bibfnamefont {Roman}\ \bibnamefont
  {Schmied}}, \ and\ \bibinfo {author} {\bibfnamefont {Philipp}\ \bibnamefont
  {Treutlein}},\ }\bibfield  {title} {\enquote {\bibinfo {title} {Quantum
  metrology with nonclassical states of atomic ensembles},}\ }\href {\doibase
  10.1103/RevModPhys.90.035005} {\bibfield  {journal} {\bibinfo  {journal}
  {Rev. Mod. Phys.}\ }\textbf {\bibinfo {volume} {90}},\ \bibinfo {pages}
  {035005} (\bibinfo {year} {2018})}\BibitemShut {NoStop}%
\bibitem [{\citenamefont {Helstrom}(1976)}]{683779922}%
  \BibitemOpen
  \bibfield  {author} {\bibinfo {author} {\bibfnamefont {C.~W.}\ \bibnamefont
  {Helstrom}},\ }\href@noop {} {\bibfield  {journal} {\bibinfo  {journal}
  {Quantum Detection and Estimation Theory}\ } (\bibinfo {year}
  {1976})}\BibitemShut {NoStop}%
\bibitem [{\citenamefont {Holevo}(1982)}]{683779921}%
  \BibitemOpen
  \bibfield  {author} {\bibinfo {author} {\bibfnamefont {A.~S.}\ \bibnamefont
  {Holevo}},\ }\href@noop {} {\bibfield  {journal} {\bibinfo  {journal}
  {Probabilistic and Statistical Aspects of Quantum Theory}\ } (\bibinfo {year}
  {1982})}\BibitemShut {NoStop}%
\bibitem [{\citenamefont {Liu}\ \emph {et~al.}(2019)\citenamefont {Liu},
  \citenamefont {Yuan}, \citenamefont {Lu},\ and\ \citenamefont
  {Wang}}]{Liu_2019}%
  \BibitemOpen
  \bibfield  {author} {\bibinfo {author} {\bibfnamefont {Jing}\ \bibnamefont
  {Liu}}, \bibinfo {author} {\bibfnamefont {Haidong}\ \bibnamefont {Yuan}},
  \bibinfo {author} {\bibfnamefont {Xiao-Ming}\ \bibnamefont {Lu}}, \ and\
  \bibinfo {author} {\bibfnamefont {Xiaoguang}\ \bibnamefont {Wang}},\
  }\bibfield  {title} {\enquote {\bibinfo {title} {Quantum fisher information
  matrix and multiparameter estimation},}\ }\href {\doibase
  10.1088/1751-8121/ab5d4d} {\bibfield  {journal} {\bibinfo  {journal} {Journal
  of Physics A: Mathematical and Theoretical}\ }\textbf {\bibinfo {volume}
  {53}},\ \bibinfo {pages} {023001} (\bibinfo {year} {2019})}\BibitemShut
  {NoStop}%
\bibitem [{\citenamefont {Ma}\ and\ \citenamefont
  {Wang}(2009)}]{PhysRevA.80.012318}%
  \BibitemOpen
  \bibfield  {author} {\bibinfo {author} {\bibfnamefont {Jian}\ \bibnamefont
  {Ma}}\ and\ \bibinfo {author} {\bibfnamefont {Xiaoguang}\ \bibnamefont
  {Wang}},\ }\bibfield  {title} {\enquote {\bibinfo {title} {Fisher information
  and spin squeezing in the lipkin-meshkov-glick model},}\ }\href {\doibase
  10.1103/PhysRevA.80.012318} {\bibfield  {journal} {\bibinfo  {journal} {Phys.
  Rev. A}\ }\textbf {\bibinfo {volume} {80}},\ \bibinfo {pages} {012318}
  (\bibinfo {year} {2009})}\BibitemShut {NoStop}%
\bibitem [{\citenamefont {Sun}\ \emph {et~al.}(2010)\citenamefont {Sun},
  \citenamefont {Ma}, \citenamefont {Lu},\ and\ \citenamefont
  {Wang}}]{PhysRevA.82.022306}%
  \BibitemOpen
  \bibfield  {author} {\bibinfo {author} {\bibfnamefont {Zhe}\ \bibnamefont
  {Sun}}, \bibinfo {author} {\bibfnamefont {Jian}\ \bibnamefont {Ma}}, \bibinfo
  {author} {\bibfnamefont {Xiao-Ming}\ \bibnamefont {Lu}}, \ and\ \bibinfo
  {author} {\bibfnamefont {Xiaoguang}\ \bibnamefont {Wang}},\ }\bibfield
  {title} {\enquote {\bibinfo {title} {Fisher information in a quantum-critical
  environment},}\ }\href {\doibase 10.1103/PhysRevA.82.022306} {\bibfield
  {journal} {\bibinfo  {journal} {Phys. Rev. A}\ }\textbf {\bibinfo {volume}
  {82}},\ \bibinfo {pages} {022306} (\bibinfo {year} {2010})}\BibitemShut
  {NoStop}%
\bibitem [{\citenamefont {Wu}\ and\ \citenamefont {Xu}(2016)}]{Wu2016}%
  \BibitemOpen
  \bibfield  {author} {\bibinfo {author} {\bibfnamefont {Wei}\ \bibnamefont
  {Wu}}\ and\ \bibinfo {author} {\bibfnamefont {Jing-Bo}\ \bibnamefont {Xu}},\
  }\bibfield  {title} {\enquote {\bibinfo {title} {Geometric phase, quantum
  fisher information, geometric quantum correlation and quantum phase
  transition in the cavity-bose--einstein-condensate system},}\ }\href
  {\doibase 10.1007/s11128-015-1186-7} {\bibfield  {journal} {\bibinfo
  {journal} {Quantum Information Processing}\ }\textbf {\bibinfo {volume}
  {15}},\ \bibinfo {pages} {3695--3709} (\bibinfo {year} {2016})}\BibitemShut
  {NoStop}%
\bibitem [{\citenamefont {Wang}\ \emph {et~al.}(2014)\citenamefont {Wang},
  \citenamefont {Wu}, \citenamefont {Yang}, \citenamefont {Jin}, \citenamefont
  {Lambert},\ and\ \citenamefont {Nori}}]{Wang_2014}%
  \BibitemOpen
  \bibfield  {author} {\bibinfo {author} {\bibfnamefont {Teng-Long}\
  \bibnamefont {Wang}}, \bibinfo {author} {\bibfnamefont {Ling-Na}\
  \bibnamefont {Wu}}, \bibinfo {author} {\bibfnamefont {Wen}\ \bibnamefont
  {Yang}}, \bibinfo {author} {\bibfnamefont {Guang-Ri}\ \bibnamefont {Jin}},
  \bibinfo {author} {\bibfnamefont {Neill}\ \bibnamefont {Lambert}}, \ and\
  \bibinfo {author} {\bibfnamefont {Franco}\ \bibnamefont {Nori}},\ }\bibfield
  {title} {\enquote {\bibinfo {title} {Quantum fisher information as a
  signature of the superradiant quantum phase transition},}\ }\href {\doibase
  10.1088/1367-2630/16/6/063039} {\bibfield  {journal} {\bibinfo  {journal}
  {New Journal of Physics}\ }\textbf {\bibinfo {volume} {16}},\ \bibinfo
  {pages} {063039} (\bibinfo {year} {2014})}\BibitemShut {NoStop}%
\bibitem [{\citenamefont {Fr\"owis}(2012)}]{PhysRevA.85.052127}%
  \BibitemOpen
  \bibfield  {author} {\bibinfo {author} {\bibfnamefont {F.}~\bibnamefont
  {Fr\"owis}},\ }\bibfield  {title} {\enquote {\bibinfo {title} {Kind of
  entanglement that speeds up quantum evolution},}\ }\href {\doibase
  10.1103/PhysRevA.85.052127} {\bibfield  {journal} {\bibinfo  {journal} {Phys.
  Rev. A}\ }\textbf {\bibinfo {volume} {85}},\ \bibinfo {pages} {052127}
  (\bibinfo {year} {2012})}\BibitemShut {NoStop}%
\bibitem [{\citenamefont {Taddei}\ \emph {et~al.}(2013)\citenamefont {Taddei},
  \citenamefont {Escher}, \citenamefont {Davidovich},\ and\ \citenamefont
  {de~Matos~Filho}}]{PhysRevLett.110.050402}%
  \BibitemOpen
  \bibfield  {author} {\bibinfo {author} {\bibfnamefont {M.~M.}\ \bibnamefont
  {Taddei}}, \bibinfo {author} {\bibfnamefont {B.~M.}\ \bibnamefont {Escher}},
  \bibinfo {author} {\bibfnamefont {L.}~\bibnamefont {Davidovich}}, \ and\
  \bibinfo {author} {\bibfnamefont {R.~L.}\ \bibnamefont {de~Matos~Filho}},\
  }\bibfield  {title} {\enquote {\bibinfo {title} {Quantum speed limit for
  physical processes},}\ }\href {\doibase 10.1103/PhysRevLett.110.050402}
  {\bibfield  {journal} {\bibinfo  {journal} {Phys. Rev. Lett.}\ }\textbf
  {\bibinfo {volume} {110}},\ \bibinfo {pages} {050402} (\bibinfo {year}
  {2013})}\BibitemShut {NoStop}%
\bibitem [{\citenamefont {Deffner}\ and\ \citenamefont
  {Campbell}(2017)}]{Deffner_2017}%
  \BibitemOpen
  \bibfield  {author} {\bibinfo {author} {\bibfnamefont {Sebastian}\
  \bibnamefont {Deffner}}\ and\ \bibinfo {author} {\bibfnamefont {Steve}\
  \bibnamefont {Campbell}},\ }\bibfield  {title} {\enquote {\bibinfo {title}
  {Quantum speed limits: from heisenberg's uncertainty principle to optimal
  quantum control},}\ }\href {\doibase 10.1088/1751-8121/aa86c6} {\bibfield
  {journal} {\bibinfo  {journal} {Journal of Physics A: Mathematical and
  Theoretical}\ }\textbf {\bibinfo {volume} {50}},\ \bibinfo {pages} {453001}
  (\bibinfo {year} {2017})}\BibitemShut {NoStop}%
\bibitem [{\citenamefont {Lu}\ \emph {et~al.}(2010)\citenamefont {Lu},
  \citenamefont {Wang},\ and\ \citenamefont {Sun}}]{PhysRevA.82.042103}%
  \BibitemOpen
  \bibfield  {author} {\bibinfo {author} {\bibfnamefont {Xiao-Ming}\
  \bibnamefont {Lu}}, \bibinfo {author} {\bibfnamefont {Xiaoguang}\
  \bibnamefont {Wang}}, \ and\ \bibinfo {author} {\bibfnamefont {C.~P.}\
  \bibnamefont {Sun}},\ }\bibfield  {title} {\enquote {\bibinfo {title}
  {Quantum fisher information flow and non-markovian processes of open
  systems},}\ }\href {\doibase 10.1103/PhysRevA.82.042103} {\bibfield
  {journal} {\bibinfo  {journal} {Phys. Rev. A}\ }\textbf {\bibinfo {volume}
  {82}},\ \bibinfo {pages} {042103} (\bibinfo {year} {2010})}\BibitemShut
  {NoStop}%
\bibitem [{\citenamefont {Song}\ \emph {et~al.}(2015)\citenamefont {Song},
  \citenamefont {Luo},\ and\ \citenamefont {Hong}}]{PhysRevA.91.042110}%
  \BibitemOpen
  \bibfield  {author} {\bibinfo {author} {\bibfnamefont {Hongting}\
  \bibnamefont {Song}}, \bibinfo {author} {\bibfnamefont {Shunlong}\
  \bibnamefont {Luo}}, \ and\ \bibinfo {author} {\bibfnamefont {Yan}\
  \bibnamefont {Hong}},\ }\bibfield  {title} {\enquote {\bibinfo {title}
  {Quantum non-markovianity based on the fisher-information matrix},}\ }\href
  {\doibase 10.1103/PhysRevA.91.042110} {\bibfield  {journal} {\bibinfo
  {journal} {Phys. Rev. A}\ }\textbf {\bibinfo {volume} {91}},\ \bibinfo
  {pages} {042110} (\bibinfo {year} {2015})}\BibitemShut {NoStop}%
\bibitem [{\citenamefont {Li}\ \emph {et~al.}(2019)\citenamefont {Li},
  \citenamefont {Guo},\ and\ \citenamefont {Piilo}}]{Li_2019}%
  \BibitemOpen
  \bibfield  {author} {\bibinfo {author} {\bibfnamefont {C.-F.}\ \bibnamefont
  {Li}}, \bibinfo {author} {\bibfnamefont {G.-C.}\ \bibnamefont {Guo}}, \ and\
  \bibinfo {author} {\bibfnamefont {J.}~\bibnamefont {Piilo}},\ }\bibfield
  {title} {\enquote {\bibinfo {title} {Non-markovian quantum dynamics: What
  does it mean?}}\ }\href {\doibase 10.1209/0295-5075/127/50001} {\bibfield
  {journal} {\bibinfo  {journal} {Europhysics Letters}\ }\textbf {\bibinfo
  {volume} {127}},\ \bibinfo {pages} {50001} (\bibinfo {year}
  {2019})}\BibitemShut {NoStop}%
\bibitem [{\citenamefont {Leggett}\ \emph {et~al.}(1987)\citenamefont
  {Leggett}, \citenamefont {Chakravarty}, \citenamefont {Dorsey}, \citenamefont
  {Fisher}, \citenamefont {Garg},\ and\ \citenamefont
  {Zwerger}}]{RevModPhys.59.1}%
  \BibitemOpen
  \bibfield  {author} {\bibinfo {author} {\bibfnamefont {A.~J.}\ \bibnamefont
  {Leggett}}, \bibinfo {author} {\bibfnamefont {S.}~\bibnamefont
  {Chakravarty}}, \bibinfo {author} {\bibfnamefont {A.~T.}\ \bibnamefont
  {Dorsey}}, \bibinfo {author} {\bibfnamefont {Matthew P.~A.}\ \bibnamefont
  {Fisher}}, \bibinfo {author} {\bibfnamefont {Anupam}\ \bibnamefont {Garg}}, \
  and\ \bibinfo {author} {\bibfnamefont {W.}~\bibnamefont {Zwerger}},\
  }\bibfield  {title} {\enquote {\bibinfo {title} {Dynamics of the dissipative
  two-state system},}\ }\href {\doibase 10.1103/RevModPhys.59.1} {\bibfield
  {journal} {\bibinfo  {journal} {Rev. Mod. Phys.}\ }\textbf {\bibinfo {volume}
  {59}},\ \bibinfo {pages} {1--85} (\bibinfo {year} {1987})}\BibitemShut
  {NoStop}%
\bibitem [{\citenamefont {Breuer}\ \emph {et~al.}(2016)\citenamefont {Breuer},
  \citenamefont {Laine}, \citenamefont {Piilo},\ and\ \citenamefont
  {Vacchini}}]{RevModPhys.88.021002}%
  \BibitemOpen
  \bibfield  {author} {\bibinfo {author} {\bibfnamefont {Heinz-Peter}\
  \bibnamefont {Breuer}}, \bibinfo {author} {\bibfnamefont {Elsi-Mari}\
  \bibnamefont {Laine}}, \bibinfo {author} {\bibfnamefont {Jyrki}\ \bibnamefont
  {Piilo}}, \ and\ \bibinfo {author} {\bibfnamefont {Bassano}\ \bibnamefont
  {Vacchini}},\ }\bibfield  {title} {\enquote {\bibinfo {title} {Colloquium:
  Non-markovian dynamics in open quantum systems},}\ }\href {\doibase
  10.1103/RevModPhys.88.021002} {\bibfield  {journal} {\bibinfo  {journal}
  {Rev. Mod. Phys.}\ }\textbf {\bibinfo {volume} {88}},\ \bibinfo {pages}
  {021002} (\bibinfo {year} {2016})}\BibitemShut {NoStop}%
\bibitem [{\citenamefont {de~Vega}\ and\ \citenamefont
  {Alonso}(2017)}]{RevModPhys.89.015001}%
  \BibitemOpen
  \bibfield  {author} {\bibinfo {author} {\bibfnamefont {In\'es}\ \bibnamefont
  {de~Vega}}\ and\ \bibinfo {author} {\bibfnamefont {Daniel}\ \bibnamefont
  {Alonso}},\ }\bibfield  {title} {\enquote {\bibinfo {title} {Dynamics of
  non-markovian open quantum systems},}\ }\href {\doibase
  10.1103/RevModPhys.89.015001} {\bibfield  {journal} {\bibinfo  {journal}
  {Rev. Mod. Phys.}\ }\textbf {\bibinfo {volume} {89}},\ \bibinfo {pages}
  {015001} (\bibinfo {year} {2017})}\BibitemShut {NoStop}%
\bibitem [{\citenamefont {Demkowicz-Dobrzanski}\ \emph
  {et~al.}(2012)\citenamefont {Demkowicz-Dobrzanski}, \citenamefont
  {Kolodynski},\ and\ \citenamefont {Guta}}]{Demkowicz2012}%
  \BibitemOpen
  \bibfield  {author} {\bibinfo {author} {\bibfnamefont {R.}~\bibnamefont
  {Demkowicz-Dobrzanski}}, \bibinfo {author} {\bibfnamefont {J.}~\bibnamefont
  {Kolodynski}}, \ and\ \bibinfo {author} {\bibfnamefont {M.}~\bibnamefont
  {Guta}},\ }\bibfield  {title} {\enquote {\bibinfo {title} {The elusive
  heisenberg limit in quantum-enhanced metrology},}\ }\href {\doibase
  10.1038/ncomms2067} {\bibfield  {journal} {\bibinfo  {journal} {Nature
  Communications}\ }\textbf {\bibinfo {volume} {3}},\ \bibinfo {pages} {1063}
  (\bibinfo {year} {2012})}\BibitemShut {NoStop}%
\bibitem [{\citenamefont {Demkowicz-Dobrza\ifmmode~\acute{n}\else
  \'{n}\fi{}ski}\ and\ \citenamefont {Maccone}(2014)}]{PhysRevLett.113.250801}%
  \BibitemOpen
  \bibfield  {author} {\bibinfo {author} {\bibfnamefont {Rafal}\ \bibnamefont
  {Demkowicz-Dobrza\ifmmode~\acute{n}\else \'{n}\fi{}ski}}\ and\ \bibinfo
  {author} {\bibfnamefont {Lorenzo}\ \bibnamefont {Maccone}},\ }\bibfield
  {title} {\enquote {\bibinfo {title} {Using entanglement against noise in
  quantum metrology},}\ }\href {\doibase 10.1103/PhysRevLett.113.250801}
  {\bibfield  {journal} {\bibinfo  {journal} {Phys. Rev. Lett.}\ }\textbf
  {\bibinfo {volume} {113}},\ \bibinfo {pages} {250801} (\bibinfo {year}
  {2014})}\BibitemShut {NoStop}%
\bibitem [{\citenamefont {Alipour}\ \emph {et~al.}(2014)\citenamefont
  {Alipour}, \citenamefont {Mehboudi},\ and\ \citenamefont
  {Rezakhani}}]{PhysRevLett.112.120405}%
  \BibitemOpen
  \bibfield  {author} {\bibinfo {author} {\bibfnamefont {S.}~\bibnamefont
  {Alipour}}, \bibinfo {author} {\bibfnamefont {M.}~\bibnamefont {Mehboudi}}, \
  and\ \bibinfo {author} {\bibfnamefont {A.~T.}\ \bibnamefont {Rezakhani}},\
  }\bibfield  {title} {\enquote {\bibinfo {title} {Quantum metrology in open
  systems: Dissipative cram\'er-rao bound},}\ }\href {\doibase
  10.1103/PhysRevLett.112.120405} {\bibfield  {journal} {\bibinfo  {journal}
  {Phys. Rev. Lett.}\ }\textbf {\bibinfo {volume} {112}},\ \bibinfo {pages}
  {120405} (\bibinfo {year} {2014})}\BibitemShut {NoStop}%
\bibitem [{\citenamefont {Mirkin}\ \emph {et~al.}(2019)\citenamefont {Mirkin},
  \citenamefont {Larocca},\ and\ \citenamefont {Wisniacki}}]{1912.04675}%
  \BibitemOpen
  \bibfield  {author} {\bibinfo {author} {\bibfnamefont {Nicolás}\
  \bibnamefont {Mirkin}}, \bibinfo {author} {\bibfnamefont {Martin}\
  \bibnamefont {Larocca}}, \ and\ \bibinfo {author} {\bibfnamefont {Diego}\
  \bibnamefont {Wisniacki}},\ }\href@noop {} {\enquote {\bibinfo {title}
  {Quantum metrology in a non-markovian quantum evolution},}\ } (\bibinfo
  {year} {2019}),\ \Eprint {http://arxiv.org/abs/arXiv:1912.04675}
  {arXiv:1912.04675} \BibitemShut {NoStop}%
\bibitem [{\citenamefont {Chin}\ \emph {et~al.}(2012)\citenamefont {Chin},
  \citenamefont {Huelga},\ and\ \citenamefont
  {Plenio}}]{PhysRevLett.109.233601}%
  \BibitemOpen
  \bibfield  {author} {\bibinfo {author} {\bibfnamefont {Alex~W.}\ \bibnamefont
  {Chin}}, \bibinfo {author} {\bibfnamefont {Susana~F.}\ \bibnamefont
  {Huelga}}, \ and\ \bibinfo {author} {\bibfnamefont {Martin~B.}\ \bibnamefont
  {Plenio}},\ }\bibfield  {title} {\enquote {\bibinfo {title} {Quantum
  metrology in non-markovian environments},}\ }\href {\doibase
  10.1103/PhysRevLett.109.233601} {\bibfield  {journal} {\bibinfo  {journal}
  {Phys. Rev. Lett.}\ }\textbf {\bibinfo {volume} {109}},\ \bibinfo {pages}
  {233601} (\bibinfo {year} {2012})}\BibitemShut {NoStop}%
\bibitem [{\citenamefont {Razavian}\ \emph {et~al.}(2019)\citenamefont
  {Razavian}, \citenamefont {Benedetti}, \citenamefont {Bina}, \citenamefont
  {Akbari-Kourbolagh},\ and\ \citenamefont {Paris}}]{Razavian2019}%
  \BibitemOpen
  \bibfield  {author} {\bibinfo {author} {\bibfnamefont {Sholeh}\ \bibnamefont
  {Razavian}}, \bibinfo {author} {\bibfnamefont {Claudia}\ \bibnamefont
  {Benedetti}}, \bibinfo {author} {\bibfnamefont {Matteo}\ \bibnamefont
  {Bina}}, \bibinfo {author} {\bibfnamefont {Yahya}\ \bibnamefont
  {Akbari-Kourbolagh}}, \ and\ \bibinfo {author} {\bibfnamefont {Matteo G.~A.}\
  \bibnamefont {Paris}},\ }\bibfield  {title} {\enquote {\bibinfo {title}
  {Quantum thermometry by single-qubit dephasing},}\ }\href {\doibase
  10.1140/epjp/i2019-12708-9} {\bibfield  {journal} {\bibinfo  {journal} {The
  European Physical Journal Plus}\ }\textbf {\bibinfo {volume} {134}},\
  \bibinfo {pages} {284} (\bibinfo {year} {2019})}\BibitemShut {NoStop}%
\bibitem [{\citenamefont {Berrada}(2013)}]{PhysRevA.88.035806}%
  \BibitemOpen
  \bibfield  {author} {\bibinfo {author} {\bibfnamefont {K.}~\bibnamefont
  {Berrada}},\ }\bibfield  {title} {\enquote {\bibinfo {title} {Non-markovian
  effect on the precision of parameter estimation},}\ }\href {\doibase
  10.1103/PhysRevA.88.035806} {\bibfield  {journal} {\bibinfo  {journal} {Phys.
  Rev. A}\ }\textbf {\bibinfo {volume} {88}},\ \bibinfo {pages} {035806}
  (\bibinfo {year} {2013})}\BibitemShut {NoStop}%
\bibitem [{\citenamefont {Tan}\ \emph {et~al.}(2013)\citenamefont {Tan},
  \citenamefont {Huang}, \citenamefont {Yin}, \citenamefont {Kuang},\ and\
  \citenamefont {Wang}}]{PhysRevA.87.032102}%
  \BibitemOpen
  \bibfield  {author} {\bibinfo {author} {\bibfnamefont {Qing-Shou}\
  \bibnamefont {Tan}}, \bibinfo {author} {\bibfnamefont {Yixiao}\ \bibnamefont
  {Huang}}, \bibinfo {author} {\bibfnamefont {Xiaolei}\ \bibnamefont {Yin}},
  \bibinfo {author} {\bibfnamefont {Le-Man}\ \bibnamefont {Kuang}}, \ and\
  \bibinfo {author} {\bibfnamefont {Xiaoguang}\ \bibnamefont {Wang}},\
  }\bibfield  {title} {\enquote {\bibinfo {title} {Enhancement of
  parameter-estimation precision in noisy systems by dynamical decoupling
  pulses},}\ }\href {\doibase 10.1103/PhysRevA.87.032102} {\bibfield  {journal}
  {\bibinfo  {journal} {Phys. Rev. A}\ }\textbf {\bibinfo {volume} {87}},\
  \bibinfo {pages} {032102} (\bibinfo {year} {2013})}\BibitemShut {NoStop}%
\bibitem [{\citenamefont {Wang}\ \emph {et~al.}(2017)\citenamefont {Wang},
  \citenamefont {Chen},\ and\ \citenamefont {An}}]{Wang_2017}%
  \BibitemOpen
  \bibfield  {author} {\bibinfo {author} {\bibfnamefont {Yuan-Sheng}\
  \bibnamefont {Wang}}, \bibinfo {author} {\bibfnamefont {Chong}\ \bibnamefont
  {Chen}}, \ and\ \bibinfo {author} {\bibfnamefont {Jun-Hong}\ \bibnamefont
  {An}},\ }\bibfield  {title} {\enquote {\bibinfo {title} {Quantum metrology in
  local dissipative environments},}\ }\href {\doibase 10.1088/1367-2630/aa8b01}
  {\bibfield  {journal} {\bibinfo  {journal} {New Journal of Physics}\ }\textbf
  {\bibinfo {volume} {19}},\ \bibinfo {pages} {113019} (\bibinfo {year}
  {2017})}\BibitemShut {NoStop}%
\bibitem [{\citenamefont {Benedetti}\ \emph {et~al.}(2018)\citenamefont
  {Benedetti}, \citenamefont {Salari~Sehdaran}, \citenamefont {Zandi},\ and\
  \citenamefont {Paris}}]{PhysRevA.97.012126}%
  \BibitemOpen
  \bibfield  {author} {\bibinfo {author} {\bibfnamefont {Claudia}\ \bibnamefont
  {Benedetti}}, \bibinfo {author} {\bibfnamefont {Fahimeh}\ \bibnamefont
  {Salari~Sehdaran}}, \bibinfo {author} {\bibfnamefont {Mohammad~H.}\
  \bibnamefont {Zandi}}, \ and\ \bibinfo {author} {\bibfnamefont {Matteo
  G.~A.}\ \bibnamefont {Paris}},\ }\bibfield  {title} {\enquote {\bibinfo
  {title} {Quantum probes for the cutoff frequency of ohmic environments},}\
  }\href {\doibase 10.1103/PhysRevA.97.012126} {\bibfield  {journal} {\bibinfo
  {journal} {Phys. Rev. A}\ }\textbf {\bibinfo {volume} {97}},\ \bibinfo
  {pages} {012126} (\bibinfo {year} {2018})}\BibitemShut {NoStop}%
\bibitem [{\citenamefont {Sehdaran}\ \emph {et~al.}(2019)\citenamefont
  {Sehdaran}, \citenamefont {Zandi},\ and\ \citenamefont
  {Bahrampour}}]{SALARISEHDARAN2019126006}%
  \BibitemOpen
  \bibfield  {author} {\bibinfo {author} {\bibfnamefont {Fahimeh~Salari}\
  \bibnamefont {Sehdaran}}, \bibinfo {author} {\bibfnamefont {Mohammad~H.}\
  \bibnamefont {Zandi}}, \ and\ \bibinfo {author} {\bibfnamefont {Alireza}\
  \bibnamefont {Bahrampour}},\ }\bibfield  {title} {\enquote {\bibinfo {title}
  {The effect of probe-ohmic environment coupling type and probe information
  flow on quantum probing of the cutoff frequency},}\ }\href {\doibase
  https://doi.org/10.1016/j.physleta.2019.126006} {\bibfield  {journal}
  {\bibinfo  {journal} {Physics Letters A}\ }\textbf {\bibinfo {volume}
  {383}},\ \bibinfo {pages} {126006} (\bibinfo {year} {2019})}\BibitemShut
  {NoStop}%
\bibitem [{\citenamefont {Tanimura}\ and\ \citenamefont
  {Kubo}(1989)}]{doi:10.1143/JPSJ.58.101}%
  \BibitemOpen
  \bibfield  {author} {\bibinfo {author} {\bibfnamefont {Yoshitaka}\
  \bibnamefont {Tanimura}}\ and\ \bibinfo {author} {\bibfnamefont {Ryogo}\
  \bibnamefont {Kubo}},\ }\bibfield  {title} {\enquote {\bibinfo {title} {Time
  evolution of a quantum system in contact with a nearly gaussian-markoffian
  noise bath},}\ }\href {\doibase 10.1143/JPSJ.58.101} {\bibfield  {journal}
  {\bibinfo  {journal} {Journal of the Physical Society of Japan}\ }\textbf
  {\bibinfo {volume} {58}},\ \bibinfo {pages} {101--114} (\bibinfo {year}
  {1989})}\BibitemShut {NoStop}%
\bibitem [{\citenamefont {an~Yan}\ \emph {et~al.}(2004)\citenamefont {an~Yan},
  \citenamefont {Yang}, \citenamefont {Liu},\ and\ \citenamefont
  {Shao}}]{YAN2004216}%
  \BibitemOpen
  \bibfield  {author} {\bibinfo {author} {\bibfnamefont {Yun}\ \bibnamefont
  {an~Yan}}, \bibinfo {author} {\bibfnamefont {Fan}\ \bibnamefont {Yang}},
  \bibinfo {author} {\bibfnamefont {Yu}~\bibnamefont {Liu}}, \ and\ \bibinfo
  {author} {\bibfnamefont {Jiushu}\ \bibnamefont {Shao}},\ }\bibfield  {title}
  {\enquote {\bibinfo {title} {Hierarchical approach based on stochastic
  decoupling to dissipative systems},}\ }\href {\doibase
  https://doi.org/10.1016/j.cplett.2004.07.036} {\bibfield  {journal} {\bibinfo
   {journal} {Chemical Physics Letters}\ }\textbf {\bibinfo {volume} {395}},\
  \bibinfo {pages} {216 -- 221} (\bibinfo {year} {2004})}\BibitemShut {NoStop}%
\bibitem [{\citenamefont {Xu}\ and\ \citenamefont
  {Yan}(2007)}]{PhysRevE.75.031107}%
  \BibitemOpen
  \bibfield  {author} {\bibinfo {author} {\bibfnamefont {Rui-Xue}\ \bibnamefont
  {Xu}}\ and\ \bibinfo {author} {\bibfnamefont {YiJing}\ \bibnamefont {Yan}},\
  }\bibfield  {title} {\enquote {\bibinfo {title} {Dynamics of quantum
  dissipation systems interacting with bosonic canonical bath: Hierarchical
  equations of motion approach},}\ }\href {\doibase 10.1103/PhysRevE.75.031107}
  {\bibfield  {journal} {\bibinfo  {journal} {Phys. Rev. E}\ }\textbf {\bibinfo
  {volume} {75}},\ \bibinfo {pages} {031107} (\bibinfo {year}
  {2007})}\BibitemShut {NoStop}%
\bibitem [{\citenamefont {Jin}\ \emph {et~al.}(2008)\citenamefont {Jin},
  \citenamefont {Zheng},\ and\ \citenamefont {Yan}}]{doi:10.1063/1.2938087}%
  \BibitemOpen
  \bibfield  {author} {\bibinfo {author} {\bibfnamefont {Jinshuang}\
  \bibnamefont {Jin}}, \bibinfo {author} {\bibfnamefont {Xiao}\ \bibnamefont
  {Zheng}}, \ and\ \bibinfo {author} {\bibfnamefont {YiJing}\ \bibnamefont
  {Yan}},\ }\bibfield  {title} {\enquote {\bibinfo {title} {Exact dynamics of
  dissipative electronic systems and quantum transport: Hierarchical equations
  of motion approach},}\ }\href {\doibase 10.1063/1.2938087} {\bibfield
  {journal} {\bibinfo  {journal} {The Journal of Chemical Physics}\ }\textbf
  {\bibinfo {volume} {128}},\ \bibinfo {pages} {234703} (\bibinfo {year}
  {2008})}\BibitemShut {NoStop}%
\bibitem [{\citenamefont {Ma}\ \emph {et~al.}(2012)\citenamefont {Ma},
  \citenamefont {Sun}, \citenamefont {Wang},\ and\ \citenamefont
  {Nori}}]{PhysRevA.85.062323}%
  \BibitemOpen
  \bibfield  {author} {\bibinfo {author} {\bibfnamefont {Jian}\ \bibnamefont
  {Ma}}, \bibinfo {author} {\bibfnamefont {Zhe}\ \bibnamefont {Sun}}, \bibinfo
  {author} {\bibfnamefont {Xiaoguang}\ \bibnamefont {Wang}}, \ and\ \bibinfo
  {author} {\bibfnamefont {Franco}\ \bibnamefont {Nori}},\ }\bibfield  {title}
  {\enquote {\bibinfo {title} {Entanglement dynamics of two qubits in a common
  bath},}\ }\href {\doibase 10.1103/PhysRevA.85.062323} {\bibfield  {journal}
  {\bibinfo  {journal} {Phys. Rev. A}\ }\textbf {\bibinfo {volume} {85}},\
  \bibinfo {pages} {062323} (\bibinfo {year} {2012})}\BibitemShut {NoStop}%
\bibitem [{\citenamefont {Wu}\ and\ \citenamefont
  {Lin}(2016)}]{PhysRevA.94.062116}%
  \BibitemOpen
  \bibfield  {author} {\bibinfo {author} {\bibfnamefont {Wei}\ \bibnamefont
  {Wu}}\ and\ \bibinfo {author} {\bibfnamefont {Hai-Qing}\ \bibnamefont
  {Lin}},\ }\bibfield  {title} {\enquote {\bibinfo {title} {Effect of bath
  temperature on the decoherence of quantum dissipative systems},}\ }\href
  {\doibase 10.1103/PhysRevA.94.062116} {\bibfield  {journal} {\bibinfo
  {journal} {Phys. Rev. A}\ }\textbf {\bibinfo {volume} {94}},\ \bibinfo
  {pages} {062116} (\bibinfo {year} {2016})}\BibitemShut {NoStop}%
\bibitem [{\citenamefont {Wu}\ and\ \citenamefont
  {Lin}(2017)}]{PhysRevA.95.042132}%
  \BibitemOpen
  \bibfield  {author} {\bibinfo {author} {\bibfnamefont {Wei}\ \bibnamefont
  {Wu}}\ and\ \bibinfo {author} {\bibfnamefont {Hai-Qing}\ \bibnamefont
  {Lin}},\ }\bibfield  {title} {\enquote {\bibinfo {title} {Quantum zeno and
  anti-zeno effects in quantum dissipative systems},}\ }\href {\doibase
  10.1103/PhysRevA.95.042132} {\bibfield  {journal} {\bibinfo  {journal} {Phys.
  Rev. A}\ }\textbf {\bibinfo {volume} {95}},\ \bibinfo {pages} {042132}
  (\bibinfo {year} {2017})}\BibitemShut {NoStop}%
\bibitem [{\citenamefont {Wu}\ and\ \citenamefont
  {Liu}(2017)}]{PhysRevA.96.032125}%
  \BibitemOpen
  \bibfield  {author} {\bibinfo {author} {\bibfnamefont {Wei}\ \bibnamefont
  {Wu}}\ and\ \bibinfo {author} {\bibfnamefont {Maoxin}\ \bibnamefont {Liu}},\
  }\bibfield  {title} {\enquote {\bibinfo {title} {Effects of
  counter-rotating-wave terms on the non-markovianity in quantum open
  systems},}\ }\href {\doibase 10.1103/PhysRevA.96.032125} {\bibfield
  {journal} {\bibinfo  {journal} {Phys. Rev. A}\ }\textbf {\bibinfo {volume}
  {96}},\ \bibinfo {pages} {032125} (\bibinfo {year} {2017})}\BibitemShut
  {NoStop}%
\bibitem [{\citenamefont {DiVincenzo}\ and\ \citenamefont
  {Loss}(2005)}]{PhysRevB.71.035318}%
  \BibitemOpen
  \bibfield  {author} {\bibinfo {author} {\bibfnamefont {David~P.}\
  \bibnamefont {DiVincenzo}}\ and\ \bibinfo {author} {\bibfnamefont {Daniel}\
  \bibnamefont {Loss}},\ }\bibfield  {title} {\enquote {\bibinfo {title}
  {Rigorous born approximation and beyond for the spin-boson model},}\ }\href
  {\doibase 10.1103/PhysRevB.71.035318} {\bibfield  {journal} {\bibinfo
  {journal} {Phys. Rev. B}\ }\textbf {\bibinfo {volume} {71}},\ \bibinfo
  {pages} {035318} (\bibinfo {year} {2005})}\BibitemShut {NoStop}%
\bibitem [{\citenamefont {Burkard}(2009)}]{PhysRevB.79.125317}%
  \BibitemOpen
  \bibfield  {author} {\bibinfo {author} {\bibfnamefont {Guido}\ \bibnamefont
  {Burkard}},\ }\bibfield  {title} {\enquote {\bibinfo {title} {Non-markovian
  qubit dynamics in the presence of $1/f$ noise},}\ }\href {\doibase
  10.1103/PhysRevB.79.125317} {\bibfield  {journal} {\bibinfo  {journal} {Phys.
  Rev. B}\ }\textbf {\bibinfo {volume} {79}},\ \bibinfo {pages} {125317}
  (\bibinfo {year} {2009})}\BibitemShut {NoStop}%
\bibitem [{\citenamefont {Huelga}\ \emph {et~al.}(1997)\citenamefont {Huelga},
  \citenamefont {Macchiavello}, \citenamefont {Pellizzari}, \citenamefont
  {Ekert}, \citenamefont {Plenio},\ and\ \citenamefont
  {Cirac}}]{PhysRevLett.79.3865}%
  \BibitemOpen
  \bibfield  {author} {\bibinfo {author} {\bibfnamefont {S.~F.}\ \bibnamefont
  {Huelga}}, \bibinfo {author} {\bibfnamefont {C.}~\bibnamefont
  {Macchiavello}}, \bibinfo {author} {\bibfnamefont {T.}~\bibnamefont
  {Pellizzari}}, \bibinfo {author} {\bibfnamefont {A.~K.}\ \bibnamefont
  {Ekert}}, \bibinfo {author} {\bibfnamefont {M.~B.}\ \bibnamefont {Plenio}}, \
  and\ \bibinfo {author} {\bibfnamefont {J.~I.}\ \bibnamefont {Cirac}},\
  }\bibfield  {title} {\enquote {\bibinfo {title} {Improvement of frequency
  standards with quantum entanglement},}\ }\href {\doibase
  10.1103/PhysRevLett.79.3865} {\bibfield  {journal} {\bibinfo  {journal}
  {Phys. Rev. Lett.}\ }\textbf {\bibinfo {volume} {79}},\ \bibinfo {pages}
  {3865--3868} (\bibinfo {year} {1997})}\BibitemShut {NoStop}%
\bibitem [{\citenamefont {Hauke}\ \emph {et~al.}(2016)\citenamefont {Hauke},
  \citenamefont {Heyl}, \citenamefont {Tagliacozzo},\ and\ \citenamefont
  {Zoller}}]{Hauke2016}%
  \BibitemOpen
  \bibfield  {author} {\bibinfo {author} {\bibfnamefont {Philipp}\ \bibnamefont
  {Hauke}}, \bibinfo {author} {\bibfnamefont {Markus}\ \bibnamefont {Heyl}},
  \bibinfo {author} {\bibfnamefont {Luca}\ \bibnamefont {Tagliacozzo}}, \ and\
  \bibinfo {author} {\bibfnamefont {Peter}\ \bibnamefont {Zoller}},\ }\bibfield
   {title} {\enquote {\bibinfo {title} {Measuring multipartite entanglement
  through dynamic susceptibilities},}\ }\href {\doibase 10.1038/nphys3700}
  {\bibfield  {journal} {\bibinfo  {journal} {Nature Physics}\ }\textbf
  {\bibinfo {volume} {12}},\ \bibinfo {pages} {778--782} (\bibinfo {year}
  {2016})}\BibitemShut {NoStop}%
\bibitem [{\citenamefont {McCormick}\ \emph {et~al.}(2019)\citenamefont
  {McCormick}, \citenamefont {Keller}, \citenamefont {Burd}, \citenamefont
  {Wineland}, \citenamefont {Wilson},\ and\ \citenamefont
  {Leibfried}}]{McCormick2019}%
  \BibitemOpen
  \bibfield  {author} {\bibinfo {author} {\bibfnamefont {Katherine~C.}\
  \bibnamefont {McCormick}}, \bibinfo {author} {\bibfnamefont {Jonas}\
  \bibnamefont {Keller}}, \bibinfo {author} {\bibfnamefont {Shaun~C.}\
  \bibnamefont {Burd}}, \bibinfo {author} {\bibfnamefont {David~J.}\
  \bibnamefont {Wineland}}, \bibinfo {author} {\bibfnamefont {Andrew~C.}\
  \bibnamefont {Wilson}}, \ and\ \bibinfo {author} {\bibfnamefont {Dietrich}\
  \bibnamefont {Leibfried}},\ }\bibfield  {title} {\enquote {\bibinfo {title}
  {Quantum-enhanced sensing of a single-ion mechanical oscillator},}\ }\href
  {\doibase 10.1038/s41586-019-1421-y} {\bibfield  {journal} {\bibinfo
  {journal} {Nature}\ }\textbf {\bibinfo {volume} {572}},\ \bibinfo {pages}
  {86--90} (\bibinfo {year} {2019})}\BibitemShut {NoStop}%
\bibitem [{\citenamefont {Haase}\ \emph {et~al.}(2018)\citenamefont {Haase},
  \citenamefont {Smirne}, \citenamefont {Ko{\l}ody{\'{n}}ski}, \citenamefont
  {Demkowicz-Dobrza{\'{n}}ski},\ and\ \citenamefont {Huelga}}]{Haase_2018}%
  \BibitemOpen
  \bibfield  {author} {\bibinfo {author} {\bibfnamefont {J~F}\ \bibnamefont
  {Haase}}, \bibinfo {author} {\bibfnamefont {A}~\bibnamefont {Smirne}},
  \bibinfo {author} {\bibfnamefont {J}~\bibnamefont {Ko{\l}ody{\'{n}}ski}},
  \bibinfo {author} {\bibfnamefont {R}~\bibnamefont
  {Demkowicz-Dobrza{\'{n}}ski}}, \ and\ \bibinfo {author} {\bibfnamefont {S~F}\
  \bibnamefont {Huelga}},\ }\bibfield  {title} {\enquote {\bibinfo {title}
  {Fundamental limits to frequency estimation: a comprehensive microscopic
  perspective},}\ }\href {\doibase 10.1088/1367-2630/aab67f} {\bibfield
  {journal} {\bibinfo  {journal} {New Journal of Physics}\ }\textbf {\bibinfo
  {volume} {20}},\ \bibinfo {pages} {053009} (\bibinfo {year}
  {2018})}\BibitemShut {NoStop}%
\bibitem [{\citenamefont {Bai}\ \emph {et~al.}(2019)\citenamefont {Bai},
  \citenamefont {Peng}, \citenamefont {Luo},\ and\ \citenamefont
  {An}}]{PhysRevLett.123.040402}%
  \BibitemOpen
  \bibfield  {author} {\bibinfo {author} {\bibfnamefont {Kai}\ \bibnamefont
  {Bai}}, \bibinfo {author} {\bibfnamefont {Zhen}\ \bibnamefont {Peng}},
  \bibinfo {author} {\bibfnamefont {Hong-Gang}\ \bibnamefont {Luo}}, \ and\
  \bibinfo {author} {\bibfnamefont {Jun-Hong}\ \bibnamefont {An}},\ }\bibfield
  {title} {\enquote {\bibinfo {title} {Retrieving ideal precision in noisy
  quantum optical metrology},}\ }\href {\doibase
  10.1103/PhysRevLett.123.040402} {\bibfield  {journal} {\bibinfo  {journal}
  {Phys. Rev. Lett.}\ }\textbf {\bibinfo {volume} {123}},\ \bibinfo {pages}
  {040402} (\bibinfo {year} {2019})}\BibitemShut {NoStop}%
\bibitem [{\citenamefont {Tamascelli}\ \emph {et~al.}(2020)\citenamefont
  {Tamascelli}, \citenamefont {Benedetti}, \citenamefont {Breuer},\ and\
  \citenamefont {Paris}}]{tamascelli2020quantum}%
  \BibitemOpen
  \bibfield  {author} {\bibinfo {author} {\bibfnamefont {Dario}\ \bibnamefont
  {Tamascelli}}, \bibinfo {author} {\bibfnamefont {Claudia}\ \bibnamefont
  {Benedetti}}, \bibinfo {author} {\bibfnamefont {Heinz-Peter}\ \bibnamefont
  {Breuer}}, \ and\ \bibinfo {author} {\bibfnamefont {Matteo G.~A.}\
  \bibnamefont {Paris}},\ }\href@noop {} {\enquote {\bibinfo {title} {Quantum
  probing beyond pure dephasing},}\ } (\bibinfo {year} {2020}),\ \Eprint
  {http://arxiv.org/abs/2003.04014} {arXiv:2003.04014 [quant-ph]} \BibitemShut
  {NoStop}%
\bibitem [{\citenamefont {Wu}(2018{\natexlab{a}})}]{PhysRevA.98.012110}%
  \BibitemOpen
  \bibfield  {author} {\bibinfo {author} {\bibfnamefont {Wei}\ \bibnamefont
  {Wu}},\ }\bibfield  {title} {\enquote {\bibinfo {title} {Realization of
  hierarchical equations of motion from stochastic perspectives},}\ }\href
  {\doibase 10.1103/PhysRevA.98.012110} {\bibfield  {journal} {\bibinfo
  {journal} {Phys. Rev. A}\ }\textbf {\bibinfo {volume} {98}},\ \bibinfo
  {pages} {012110} (\bibinfo {year} {2018}{\natexlab{a}})}\BibitemShut
  {NoStop}%
\bibitem [{\citenamefont {Wu}(2018{\natexlab{b}})}]{PhysRevA.98.032116}%
  \BibitemOpen
  \bibfield  {author} {\bibinfo {author} {\bibfnamefont {Wei}\ \bibnamefont
  {Wu}},\ }\bibfield  {title} {\enquote {\bibinfo {title} {Stochastic
  decoupling approach to the spin-boson dynamics: Perturbative and
  nonperturbative treatments},}\ }\href {\doibase 10.1103/PhysRevA.98.032116}
  {\bibfield  {journal} {\bibinfo  {journal} {Phys. Rev. A}\ }\textbf {\bibinfo
  {volume} {98}},\ \bibinfo {pages} {032116} (\bibinfo {year}
  {2018}{\natexlab{b}})}\BibitemShut {NoStop}%
\bibitem [{\citenamefont {Suess}\ \emph {et~al.}(2014)\citenamefont {Suess},
  \citenamefont {Eisfeld},\ and\ \citenamefont
  {Strunz}}]{PhysRevLett.113.150403}%
  \BibitemOpen
  \bibfield  {author} {\bibinfo {author} {\bibfnamefont {D.}~\bibnamefont
  {Suess}}, \bibinfo {author} {\bibfnamefont {A.}~\bibnamefont {Eisfeld}}, \
  and\ \bibinfo {author} {\bibfnamefont {W.~T.}\ \bibnamefont {Strunz}},\
  }\bibfield  {title} {\enquote {\bibinfo {title} {Hierarchy of stochastic pure
  states for open quantum system dynamics},}\ }\href {\doibase
  10.1103/PhysRevLett.113.150403} {\bibfield  {journal} {\bibinfo  {journal}
  {Phys. Rev. Lett.}\ }\textbf {\bibinfo {volume} {113}},\ \bibinfo {pages}
  {150403} (\bibinfo {year} {2014})}\BibitemShut {NoStop}%
\bibitem [{\citenamefont {Suess}\ \emph {et~al.}(2015)\citenamefont {Suess},
  \citenamefont {Strunz},\ and\ \citenamefont {Eisfeld}}]{Suess2015}%
  \BibitemOpen
  \bibfield  {author} {\bibinfo {author} {\bibfnamefont {D.}~\bibnamefont
  {Suess}}, \bibinfo {author} {\bibfnamefont {W.~T.}\ \bibnamefont {Strunz}}, \
  and\ \bibinfo {author} {\bibfnamefont {A.}~\bibnamefont {Eisfeld}},\
  }\bibfield  {title} {\enquote {\bibinfo {title} {Hierarchical equations for
  open system dynamics in fermionic and bosonic environments},}\ }\href
  {\doibase 10.1007/s10955-015-1236-7} {\bibfield  {journal} {\bibinfo
  {journal} {Journal of Statistical Physics}\ }\textbf {\bibinfo {volume}
  {159}},\ \bibinfo {pages} {1408--1423} (\bibinfo {year} {2015})}\BibitemShut
  {NoStop}%
\bibitem [{\citenamefont {Nakajima}(1958)}]{10.1143/PTP.20.948}%
  \BibitemOpen
  \bibfield  {author} {\bibinfo {author} {\bibfnamefont {Sadao}\ \bibnamefont
  {Nakajima}},\ }\bibfield  {title} {\enquote {\bibinfo {title} {{On Quantum
  Theory of Transport Phenomena: Steady Diffusion}},}\ }\href {\doibase
  10.1143/PTP.20.948} {\bibfield  {journal} {\bibinfo  {journal} {Progress of
  Theoretical Physics}\ }\textbf {\bibinfo {volume} {20}},\ \bibinfo {pages}
  {948--959} (\bibinfo {year} {1958})}\BibitemShut {NoStop}%
\bibitem [{\citenamefont {Zwanzig}(1960)}]{doi:10.1063/1.1731409}%
  \BibitemOpen
  \bibfield  {author} {\bibinfo {author} {\bibfnamefont {Robert}\ \bibnamefont
  {Zwanzig}},\ }\bibfield  {title} {\enquote {\bibinfo {title} {Ensemble method
  in the theory of irreversibility},}\ }\href {\doibase 10.1063/1.1731409}
  {\bibfield  {journal} {\bibinfo  {journal} {The Journal of Chemical Physics}\
  }\textbf {\bibinfo {volume} {33}},\ \bibinfo {pages} {1338--1341} (\bibinfo
  {year} {1960})}\BibitemShut {NoStop}%
\bibitem [{\citenamefont {Wu}\ and\ \citenamefont {Zhu}(2020)}]{WU2020168203}%
  \BibitemOpen
  \bibfield  {author} {\bibinfo {author} {\bibfnamefont {Wei}\ \bibnamefont
  {Wu}}\ and\ \bibinfo {author} {\bibfnamefont {Wen-Li}\ \bibnamefont {Zhu}},\
  }\bibfield  {title} {\enquote {\bibinfo {title} {Heat transfer in a
  nonequilibrium spin-boson model: A perturbative approach},}\ }\href {\doibase
  https://doi.org/10.1016/j.aop.2020.168203} {\bibfield  {journal} {\bibinfo
  {journal} {Annals of Physics}\ }\textbf {\bibinfo {volume} {418}},\ \bibinfo
  {pages} {168203} (\bibinfo {year} {2020})}\BibitemShut {NoStop}%
\bibitem [{\citenamefont {Shen}\ \emph {et~al.}(2014)\citenamefont {Shen},
  \citenamefont {Qin}, \citenamefont {Xiu},\ and\ \citenamefont
  {Yi}}]{PhysRevA.89.062113}%
  \BibitemOpen
  \bibfield  {author} {\bibinfo {author} {\bibfnamefont {H.~Z.}\ \bibnamefont
  {Shen}}, \bibinfo {author} {\bibfnamefont {M.}~\bibnamefont {Qin}}, \bibinfo
  {author} {\bibfnamefont {Xiao-Ming}\ \bibnamefont {Xiu}}, \ and\ \bibinfo
  {author} {\bibfnamefont {X.~X.}\ \bibnamefont {Yi}},\ }\bibfield  {title}
  {\enquote {\bibinfo {title} {Exact non-markovian master equation for a driven
  damped two-level system},}\ }\href {\doibase 10.1103/PhysRevA.89.062113}
  {\bibfield  {journal} {\bibinfo  {journal} {Phys. Rev. A}\ }\textbf {\bibinfo
  {volume} {89}},\ \bibinfo {pages} {062113} (\bibinfo {year}
  {2014})}\BibitemShut {NoStop}%
\bibitem [{\citenamefont {Bellomo}\ \emph {et~al.}(2007)\citenamefont
  {Bellomo}, \citenamefont {Lo~Franco},\ and\ \citenamefont
  {Compagno}}]{PhysRevLett.99.160502}%
  \BibitemOpen
  \bibfield  {author} {\bibinfo {author} {\bibfnamefont {B.}~\bibnamefont
  {Bellomo}}, \bibinfo {author} {\bibfnamefont {R.}~\bibnamefont {Lo~Franco}},
  \ and\ \bibinfo {author} {\bibfnamefont {G.}~\bibnamefont {Compagno}},\
  }\bibfield  {title} {\enquote {\bibinfo {title} {Non-markovian effects on the
  dynamics of entanglement},}\ }\href {\doibase 10.1103/PhysRevLett.99.160502}
  {\bibfield  {journal} {\bibinfo  {journal} {Phys. Rev. Lett.}\ }\textbf
  {\bibinfo {volume} {99}},\ \bibinfo {pages} {160502} (\bibinfo {year}
  {2007})}\BibitemShut {NoStop}%
\bibitem [{\citenamefont {Li}\ \emph {et~al.}(2010)\citenamefont {Li},
  \citenamefont {Zou},\ and\ \citenamefont {Shao}}]{PhysRevA.81.062124}%
  \BibitemOpen
  \bibfield  {author} {\bibinfo {author} {\bibfnamefont {Jun-Gang}\
  \bibnamefont {Li}}, \bibinfo {author} {\bibfnamefont {Jian}\ \bibnamefont
  {Zou}}, \ and\ \bibinfo {author} {\bibfnamefont {Bin}\ \bibnamefont {Shao}},\
  }\bibfield  {title} {\enquote {\bibinfo {title} {Non-markovianity of the
  damped jaynes-cummings model with detuning},}\ }\href {\doibase
  10.1103/PhysRevA.81.062124} {\bibfield  {journal} {\bibinfo  {journal} {Phys.
  Rev. A}\ }\textbf {\bibinfo {volume} {81}},\ \bibinfo {pages} {062124}
  (\bibinfo {year} {2010})}\BibitemShut {NoStop}%
\bibitem [{\citenamefont {Wu}\ and\ \citenamefont {Cheng}(2018)}]{Wu2018}%
  \BibitemOpen
  \bibfield  {author} {\bibinfo {author} {\bibfnamefont {Wei}\ \bibnamefont
  {Wu}}\ and\ \bibinfo {author} {\bibfnamefont {Jun-Qing}\ \bibnamefont
  {Cheng}},\ }\bibfield  {title} {\enquote {\bibinfo {title} {Coherent dynamics
  of a qubit--oscillator system in a noisy environment},}\ }\href {\doibase
  10.1007/s11128-018-2071-y} {\bibfield  {journal} {\bibinfo  {journal}
  {Quantum Information Processing}\ }\textbf {\bibinfo {volume} {17}},\
  \bibinfo {pages} {300} (\bibinfo {year} {2018})}\BibitemShut {NoStop}%
\bibitem [{\citenamefont {Hou}\ \emph {et~al.}(2011)\citenamefont {Hou},
  \citenamefont {Yi}, \citenamefont {Yu},\ and\ \citenamefont
  {Oh}}]{PhysRevA.83.062115}%
  \BibitemOpen
  \bibfield  {author} {\bibinfo {author} {\bibfnamefont {S.~C.}\ \bibnamefont
  {Hou}}, \bibinfo {author} {\bibfnamefont {X.~X.}\ \bibnamefont {Yi}},
  \bibinfo {author} {\bibfnamefont {S.~X.}\ \bibnamefont {Yu}}, \ and\ \bibinfo
  {author} {\bibfnamefont {C.~H.}\ \bibnamefont {Oh}},\ }\bibfield  {title}
  {\enquote {\bibinfo {title} {Alternative non-markovianity measure by
  divisibility of dynamical maps},}\ }\href {\doibase
  10.1103/PhysRevA.83.062115} {\bibfield  {journal} {\bibinfo  {journal} {Phys.
  Rev. A}\ }\textbf {\bibinfo {volume} {83}},\ \bibinfo {pages} {062115}
  (\bibinfo {year} {2011})}\BibitemShut {NoStop}%
\bibitem [{\citenamefont {Sun}\ \emph {et~al.}(2016)\citenamefont {Sun},
  \citenamefont {Fujihashi}, \citenamefont {Ishizaki},\ and\ \citenamefont
  {Zhao}}]{doi:10.1063/1.4950888}%
  \BibitemOpen
  \bibfield  {author} {\bibinfo {author} {\bibfnamefont {Ke-Wei}\ \bibnamefont
  {Sun}}, \bibinfo {author} {\bibfnamefont {Yuta}\ \bibnamefont {Fujihashi}},
  \bibinfo {author} {\bibfnamefont {Akihito}\ \bibnamefont {Ishizaki}}, \ and\
  \bibinfo {author} {\bibfnamefont {Yang}\ \bibnamefont {Zhao}},\ }\bibfield
  {title} {\enquote {\bibinfo {title} {A variational master equation approach
  to quantum dynamics with off-diagonal coupling in a sub-ohmic environment},}\
  }\href {\doibase 10.1063/1.4950888} {\bibfield  {journal} {\bibinfo
  {journal} {The Journal of Chemical Physics}\ }\textbf {\bibinfo {volume}
  {144}},\ \bibinfo {pages} {204106} (\bibinfo {year} {2016})}\BibitemShut
  {NoStop}%
\bibitem [{\citenamefont {Galperin}\ \emph {et~al.}(2006)\citenamefont
  {Galperin}, \citenamefont {Altshuler}, \citenamefont {Bergli},\ and\
  \citenamefont {Shantsev}}]{PhysRevLett.96.097009}%
  \BibitemOpen
  \bibfield  {author} {\bibinfo {author} {\bibfnamefont {Y.~M.}\ \bibnamefont
  {Galperin}}, \bibinfo {author} {\bibfnamefont {B.~L.}\ \bibnamefont
  {Altshuler}}, \bibinfo {author} {\bibfnamefont {J.}~\bibnamefont {Bergli}}, \
  and\ \bibinfo {author} {\bibfnamefont {D.~V.}\ \bibnamefont {Shantsev}},\
  }\bibfield  {title} {\enquote {\bibinfo {title} {Non-gaussian low-frequency
  noise as a source of qubit decoherence},}\ }\href {\doibase
  10.1103/PhysRevLett.96.097009} {\bibfield  {journal} {\bibinfo  {journal}
  {Phys. Rev. Lett.}\ }\textbf {\bibinfo {volume} {96}},\ \bibinfo {pages}
  {097009} (\bibinfo {year} {2006})}\BibitemShut {NoStop}%
\bibitem [{\citenamefont {Cai}\ and\ \citenamefont
  {Zheng}(2018)}]{doi:10.1063/1.5039891}%
  \BibitemOpen
  \bibfield  {author} {\bibinfo {author} {\bibfnamefont {Xiangji}\ \bibnamefont
  {Cai}}\ and\ \bibinfo {author} {\bibfnamefont {Yujun}\ \bibnamefont
  {Zheng}},\ }\bibfield  {title} {\enquote {\bibinfo {title} {Non-markovian
  decoherence dynamics in nonequilibrium environments},}\ }\href {\doibase
  10.1063/1.5039891} {\bibfield  {journal} {\bibinfo  {journal} {The Journal of
  Chemical Physics}\ }\textbf {\bibinfo {volume} {149}},\ \bibinfo {pages}
  {094107} (\bibinfo {year} {2018})}\BibitemShut {NoStop}%
\bibitem [{\citenamefont {Tang}\ \emph {et~al.}(2015)\citenamefont {Tang},
  \citenamefont {Ouyang}, \citenamefont {Gong}, \citenamefont {Wang},\ and\
  \citenamefont {Wu}}]{doi:10.1063/1.4936924}%
  \BibitemOpen
  \bibfield  {author} {\bibinfo {author} {\bibfnamefont {Zhoufei}\ \bibnamefont
  {Tang}}, \bibinfo {author} {\bibfnamefont {Xiaolong}\ \bibnamefont {Ouyang}},
  \bibinfo {author} {\bibfnamefont {Zhihao}\ \bibnamefont {Gong}}, \bibinfo
  {author} {\bibfnamefont {Haobin}\ \bibnamefont {Wang}}, \ and\ \bibinfo
  {author} {\bibfnamefont {Jianlan}\ \bibnamefont {Wu}},\ }\bibfield  {title}
  {\enquote {\bibinfo {title} {Extended hierarchy equation of motion for the
  spin-boson model},}\ }\href {\doibase 10.1063/1.4936924} {\bibfield
  {journal} {\bibinfo  {journal} {The Journal of Chemical Physics}\ }\textbf
  {\bibinfo {volume} {143}},\ \bibinfo {pages} {224112} (\bibinfo {year}
  {2015})}\BibitemShut {NoStop}%
\bibitem [{\citenamefont {Duan}\ \emph {et~al.}(2017)\citenamefont {Duan},
  \citenamefont {Tang}, \citenamefont {Cao},\ and\ \citenamefont
  {Wu}}]{PhysRevB.95.214308}%
  \BibitemOpen
  \bibfield  {author} {\bibinfo {author} {\bibfnamefont {Chenru}\ \bibnamefont
  {Duan}}, \bibinfo {author} {\bibfnamefont {Zhoufei}\ \bibnamefont {Tang}},
  \bibinfo {author} {\bibfnamefont {Jianshu}\ \bibnamefont {Cao}}, \ and\
  \bibinfo {author} {\bibfnamefont {Jianlan}\ \bibnamefont {Wu}},\ }\bibfield
  {title} {\enquote {\bibinfo {title} {Zero-temperature localization in a
  sub-ohmic spin-boson model investigated by an extended hierarchy equation of
  motion},}\ }\href {\doibase 10.1103/PhysRevB.95.214308} {\bibfield  {journal}
  {\bibinfo  {journal} {Phys. Rev. B}\ }\textbf {\bibinfo {volume} {95}},\
  \bibinfo {pages} {214308} (\bibinfo {year} {2017})}\BibitemShut {NoStop}%
\bibitem [{\citenamefont {Zhang}\ \emph {et~al.}(2020)\citenamefont {Zhang},
  \citenamefont {Cui}, \citenamefont {Gong}, \citenamefont {Xu}, \citenamefont
  {Zheng},\ and\ \citenamefont {Yan}}]{doi:10.1063/1.5136093}%
  \BibitemOpen
  \bibfield  {author} {\bibinfo {author} {\bibfnamefont {Hou-Dao}\ \bibnamefont
  {Zhang}}, \bibinfo {author} {\bibfnamefont {Lei}\ \bibnamefont {Cui}},
  \bibinfo {author} {\bibfnamefont {Hong}\ \bibnamefont {Gong}}, \bibinfo
  {author} {\bibfnamefont {Rui-Xue}\ \bibnamefont {Xu}}, \bibinfo {author}
  {\bibfnamefont {Xiao}\ \bibnamefont {Zheng}}, \ and\ \bibinfo {author}
  {\bibfnamefont {YiJing}\ \bibnamefont {Yan}},\ }\bibfield  {title} {\enquote
  {\bibinfo {title} {Hierarchical equations of motion method based on fano
  spectrum decomposition for low temperature environments},}\ }\href {\doibase
  10.1063/1.5136093} {\bibfield  {journal} {\bibinfo  {journal} {The Journal of
  Chemical Physics}\ }\textbf {\bibinfo {volume} {152}},\ \bibinfo {pages}
  {064107} (\bibinfo {year} {2020})}\BibitemShut {NoStop}%
\bibitem [{\citenamefont {Hsieh}\ and\ \citenamefont
  {Cao}(2018{\natexlab{a}})}]{doi:10.1063/1.5018725}%
  \BibitemOpen
  \bibfield  {author} {\bibinfo {author} {\bibfnamefont {Chang-Yu}\
  \bibnamefont {Hsieh}}\ and\ \bibinfo {author} {\bibfnamefont {Jianshu}\
  \bibnamefont {Cao}},\ }\bibfield  {title} {\enquote {\bibinfo {title} {A
  unified stochastic formulation of dissipative quantum dynamics. i.
  generalized hierarchical equations},}\ }\href {\doibase 10.1063/1.5018725}
  {\bibfield  {journal} {\bibinfo  {journal} {The Journal of Chemical Physics}\
  }\textbf {\bibinfo {volume} {148}},\ \bibinfo {pages} {014103} (\bibinfo
  {year} {2018}{\natexlab{a}})}\BibitemShut {NoStop}%
\bibitem [{\citenamefont {Hsieh}\ and\ \citenamefont
  {Cao}(2018{\natexlab{b}})}]{doi:10.1063/1.5018726}%
  \BibitemOpen
  \bibfield  {author} {\bibinfo {author} {\bibfnamefont {Chang-Yu}\
  \bibnamefont {Hsieh}}\ and\ \bibinfo {author} {\bibfnamefont {Jianshu}\
  \bibnamefont {Cao}},\ }\bibfield  {title} {\enquote {\bibinfo {title} {A
  unified stochastic formulation of dissipative quantum dynamics. ii. beyond
  linear response of spin baths},}\ }\href {\doibase 10.1063/1.5018726}
  {\bibfield  {journal} {\bibinfo  {journal} {The Journal of Chemical Physics}\
  }\textbf {\bibinfo {volume} {148}},\ \bibinfo {pages} {014104} (\bibinfo
  {year} {2018}{\natexlab{b}})}\BibitemShut {NoStop}%
\bibitem [{\citenamefont {Diósi}\ and\ \citenamefont
  {Strunz}(1997)}]{DIOSI1997569}%
  \BibitemOpen
  \bibfield  {author} {\bibinfo {author} {\bibfnamefont {Lajos}\ \bibnamefont
  {Diósi}}\ and\ \bibinfo {author} {\bibfnamefont {Walter~T.}\ \bibnamefont
  {Strunz}},\ }\bibfield  {title} {\enquote {\bibinfo {title} {The
  non-markovian stochastic schrödinger equation for open systems},}\ }\href
  {\doibase https://doi.org/10.1016/S0375-9601(97)00717-2} {\bibfield
  {journal} {\bibinfo  {journal} {Physics Letters A}\ }\textbf {\bibinfo
  {volume} {235}},\ \bibinfo {pages} {569 -- 573} (\bibinfo {year}
  {1997})}\BibitemShut {NoStop}%
\bibitem [{\citenamefont {Strunz}(2001)}]{STRUNZ2001237}%
  \BibitemOpen
  \bibfield  {author} {\bibinfo {author} {\bibfnamefont {Walter~T.}\
  \bibnamefont {Strunz}},\ }\bibfield  {title} {\enquote {\bibinfo {title} {The
  brownian motion stochastic schrödinger equation},}\ }\href {\doibase
  https://doi.org/10.1016/S0301-0104(01)00299-3} {\bibfield  {journal}
  {\bibinfo  {journal} {Chemical Physics}\ }\textbf {\bibinfo {volume} {268}},\
  \bibinfo {pages} {237 -- 248} (\bibinfo {year} {2001})}\BibitemShut {NoStop}%
\end{thebibliography}%

\end{document}